\documentclass[twocolumn, prx, superscriptaddress,notitlepage]{revtex4-2}
\usepackage{graphicx}
\usepackage{dcolumn}
\usepackage{bm}
\usepackage[usenames,dvipsnames]{color}
\usepackage[most]{tcolorbox}
\usepackage{multirow}
\usepackage{gensymb}
\usepackage[normalem]{ulem}
\usepackage{CJK}
\usepackage{comment}
\usepackage[colorlinks, linkcolor=blue,anchorcolor=blue,citecolor=blue,urlcolor=blue]{hyperref}
\usepackage{amssymb}
\usepackage{pifont}
\usepackage{physics}
\usepackage{natbib}
\usepackage{units} 
\usepackage{booktabs}
\usepackage{colortbl}

\usepackage{color,soul}
\usepackage[mathscr]{euscript}
\begin{document}
\begin{CJK*}{UTF8}{}
\title{Dual-type dual-element atom arrays for quantum information processing}

\author{Zhanchuan Zhang}
\affiliation{
   Department of Physics/Institute of Quantum Electronics and Quantum Center, ETH Z\"{u}rich, Z\"{u}rich 8093, Switzerland}
 \affiliation{
   Laboratory for Nano and Quantum Technologies, Paul Scherrer Institut, CH-5232 Villigen PSI, Switzerland}  

\author{Jeth Arunseangroj}
\affiliation{
   Department of Physics/Institute of Quantum Electronics and Quantum Center, ETH Z\"{u}rich, Z\"{u}rich 8093, Switzerland}
\affiliation{
   Laboratory for Nano and Quantum Technologies, Paul Scherrer Institut, CH-5232 Villigen PSI, Switzerland}  

\author{Wenchao Xu}
\affiliation{
   Department of Physics/Institute of Quantum Electronics and Quantum Center, ETH Z\"{u}rich, Z\"{u}rich 8093, Switzerland}
\affiliation{
   Laboratory for Nano and Quantum Technologies, Paul Scherrer Institut, CH-5232 Villigen PSI, Switzerland}  

\begin{abstract}

Neutral-atom arrays are a leading platform for quantum technologies, offering a promising route toward large-scale, fault-tolerant quantum computing. We propose a novel quantum processing architecture based on dual-type, dual-element atom arrays, where individually trapped atoms serve as data qubits, and small atomic ensembles enable ancillary operations. By leveraging the selective initialization, coherent control, and collective optical response of atomic ensembles, we demonstrate ensemble-assisted quantum operations that enable reconfigurable, high-speed control of individual data qubits and rapid mid-circuit readout, including both projective single-qubit and joint multi-qubit measurements. The hybrid approach of this architecture combines the long coherence times of single-atom qubits with the enhanced controllability of atomic ensembles, achieving high-fidelity state manipulation and detection with minimal crosstalk. Numerical simulations indicate that our scheme supports individually addressable single- and multi-qubit operations with fidelities of 99.5\% and 99.9\%, respectively, as well as fast single- and multi-qubit state readout with fidelities exceeding 99\% within tens of microseconds. These capabilities open new pathways toward scalable, fault-tolerant quantum computation, enabling repetitive error syndrome detection and efficient generation of long-range entangled many-body states, thereby expanding the quantum information toolbox beyond existing platforms.

\end{abstract}

\maketitle
\end{CJK*}




\section{Introduction}
Arrays of neutral-atom qubits, where each atom is individually trapped atoms and controlled, have emerged as a highly promising platform for programmable quantum processors, enabling both digital quantum computation and analog quantum simulation~\cite{henriet2020quantum, morgado2021quantum, adams2019rydberg, wu2021concise}. This architecture offers remarkable scalability, controllability, and programmability, making it suitable for advancing quantum technologies. Notable achievements in this field include the simulation of quantum spin models~\cite{Bernien2017probing, Ebadi2021quantum, Scholl2021quantum, semeghini2021probing, chen2023continuous, zhao2025observation, manovitz2025quantum}, the implementation of high-fidelity quantum gate operations~\cite{levine2019parallel, Evered2023high-fidelity,Graham2022multi-qubit}, and advancements in quantum-enhanced metrology~\cite{Eckner2023realizing, Bornet2023scalable, cao2024multiqubit, shaw2024multi, finkelstein2024universal}. Remarkably, recent breakthroughs have demonstrated the implementation of high-fidelity, error-correctable quantum gates, the execution of complex sampling circuits with 48 logical qubits based on 280 physical qubits~\cite{Bluvstein2024logical}, and magic state distillation using logical qubits~\cite{rodriguez2024experimental}, highlighting the rapid progress of fault-tolerant quantum computation. 




Despite these advancements, further development of this platform faces key challenges that require improvement, including the necessity for reconfigurable individual addressability in qubit/spin operations and non-demolition detection of a subset of atoms without crosstalk. These limitations reduce the efficiency of quantum algorithms, hinder experimental repetition rates, and restrict the implementation of many measurement-based protocols and quantum error correction schemes. A promising approach capable of mitigating these challenges is a hybrid quantum processing architecture, which comprises more than one type of atomic element. Such a system provides greater flexibility in controlling both interspecies and intra-species interactions, as well as allows the development of crosstalk-free protocols for mid-circuit measurements. Recent experimental advancements have successfully demonstrated the creation of large-size dual-element atom arrays, achieving independent cooling and loading of different atomic species into arbitrarily configurable two-dimensional optical tweezer arrays~\cite{singh2022dual-element, Sheng2022defect, nakamura2024hybrid, wei2025enhanced}. Furthermore, controlled quantum gates and quantum state transfer between different atomic species have also been experimentally realized~\cite{anand2024dualspecies}. However, whether this approach represents the optimal balance between technical overhead and functionality remains an open question, motivating the exploration of alternative strategies. 

In this work, we propose a new quantum architecture that combines different types of atomic arrays to address current challenges while maintaining minimal technical overhead. This dual-type atom array integrates individually trapped atoms with atomic ensembles, leveraging the unique strengths of each system for quantum information processing. Quantum information is encoded in the long-lived ground states of individual atoms, while atomic ensembles of a different atomic species facilitate ancillary operations on the data qubits. By exploiting controllable interactions between single atoms and atomic ensembles via their Rydberg states, as well as the strong collective optical effects of ensembles, we develop a versatile toolbox for reconfigurable individual quantum control and rapid non-demolition detection of single- and multi-qubit states. This enhanced quantum control, combined with the significantly reduced measurement times and compatibility with continuous operation modes, largely improves the efficiency of implementing measurement-based quantum processing protocols and fault-tolerant quantum computation. 





\section{Overview of dual-type atom array architecture}
\label{Experimental implementation}

\begin{figure}
    \centering
    \includegraphics[width=\linewidth]{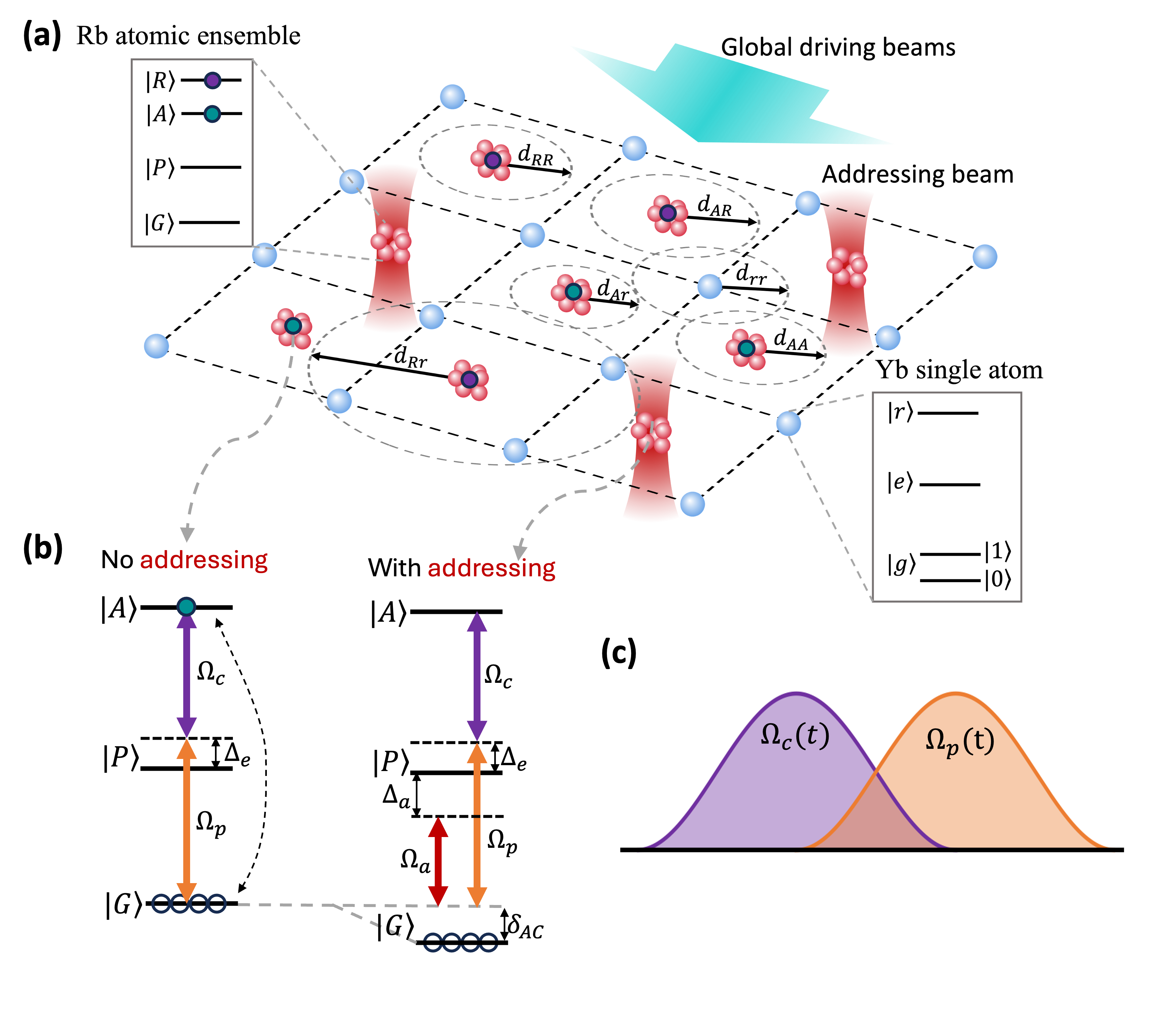}
    \caption{(a): Schematic of the dual-type dual-element quantum processing architecture, consisting of individually trapped $^{171}$Yb atoms as data qubits (blue spheres) and atomic ensembles of $^{87}$Rb for ancillary operations (red clouds). The relevant energy levels of atoms are shown. To selectively prepare a target set of atomic ensembles in the collective excited state $\ket{A}$ (green dots), global Rydberg excitation lasers are combined with an additional set of trapping light beams for local addressing. $\ket{A}$ can be coherently coupled to another Rydberg state $\ket{R}$ (purple dots) via microwave transitions. Controllable strong interactions between individual Yb atoms and excited atomic ensembles through their Rydberg states $\ket{R}$ and $\ket{r}$ enable ensemble-assisted ancillary operations with rapid, reconfigurable local addressability and minimal crosstalk. Mobile optical tweezers allow shuttling of Rb atomic ensembles and dynamic adjustment of the geometrical configuration for various functionalities. The dashed circles show the Rydberg blockade radius for different Rydberg state pairs, with only $d_{Rr}$ larger than the inter-atomic distance. (b): Local-addressing beams enable site-selective control of atomic ensemble states. Initially, all atoms are in the ground state. Through proper temporal control of the global excitation beams $\Omega_p(t)$ and $\Omega_c(t)$, the atomic ensemble adiabatically evolves into $\ket{A}$ in the absence of addressing beam. When addressing beams are applied, they locally shift the energy separation between $\ket{G}$ and $\ket{A}$, preventing Rydberg excitation and enabling site-selective operations. (c): The time-dependent profiles of the global excitation beams $\Omega_p(t)$ and $\Omega_c(t)$. These beams adiabatically initialize atomic ensembles into state $\ket{A}$.}
    \label{architecture}
\end{figure}
Fig.~\ref{architecture}(a) depicts the experimental realization of the proposed quantum processing architecture, featuring two types of atom arrays composed of different atomic species. The natural difference in transition frequencies of the two atomic species allows independent control of the two atom arrays and provides the flexibility to dynamically reconfigure their relative positions based on the specific experimental objectives. Unlike other existing dual-species arrays, here, each optical tweezer in the second array hosts a small atomic ensemble containing a few hundred atoms. The radius of each atomic ensemble is kept small (on the order of a few $\mu$m), ensuring that at most a single Rydberg excitation can be created within the ensemble, due to the Rydberg blockade effect~\cite{Urban2009observation, Gaetan2009observation}. Leveraging the simplicity of rapid control over the atomic ensembles' states and their strong optical nonlinearity, here, we introduce novel schemes in which atomic ensembles facilitate ancillary operations on data qubits. 



In this work, we focus on a specific choice of atomic species as the building blocks for dual-type dual-element atom arrays. The array of individually trapped atoms is composed of $^{171}$Yb atoms, generated using holographic imaging and rearrangement techniques~\cite{endres2016atom-by-atom, barredo2016atom-by-atom, Kim2016insitu}. As an alkaline-earth-like atomic species with a nuclear spin of $I=1/2$ and a rich energy structure, $^{171}$Yb has emerged as a promising building block for quantum technologies. This system offers a long coherence time for storing quantum information within its nuclear spin degrees of freedom, enabling high-fidelity quantum gates, and supporting versatile encoding schemes, including erasure error correction~\cite{ma2022universal, jenkins2022ytterbium, Wu2022erasure, Ma2023high-fidelity, lis2023midcircuit, jia2024architecture, Scholl2023erasure, peper2025spectroscopy, muniz2024high, reichardt2024logical}.
On the other hand, the array of atomic ensembles consists of $^{87}$Rb atoms, which are technically simple for their well-established laser cooling and trapping techniques and having extensively studied Rydberg properties. The generation of large, defect-free arrays of small atomic ensembles can be efficiently achieved using commercially available optical modulators, without the need for atom rearrangement~\cite{Wang2020preparation}. While our discussion primarily focuses on Yb-Rb dual-type arrays, the proposed schemes can be readily extended to other atomic species with minimal modifications (see SM for further discussion).

Within this dual-type array, quantum information is encoded in the long-lived nuclear spin states of individually trapped Yb atoms, with $\ket{0}\equiv \ket{^1S_0,m_F=-1/2}$ and $\ket{1}\equiv \ket{^1S_0,m_F=1/2}$. Ancillary operations on Yb atoms are performed using Rb atomic ensembles. Most of the time, these ensembles remain in their ground state, denoted as $\ket{G}=\ket{gg...g}$, where each atom in the ensemble is at state $\ket{g}\equiv \ket{F=2,m_F=2}$. Ground-state atoms can be excited to a Rydberg state $\ket{A}$ via two-photon transitions. In sufficiently small-size atomic ensembles, strong Rydberg blockade constrains the Hilbert space, effectively reducing the N-atom ensemble to a two-level system. We denote the excited state of this collective system as $\ket{A}$, where a Rydberg excitation is coherently shared among the atoms in the ensemble. To implement our proposed protocols, an additional Rydberg state $\ket{R}$ is introduced, which is coherently coupled to $\ket{A}$ via global microwave driving fields. 


Coherent manipulation of the states in a Rydberg-blockaded atomic ensemble requires careful consideration due to the ensemble's unique properties compared to single atoms. The Rabi coupling between $\ket{G}$ and $\ket{A}$ is collectively enhanced, scaling as $\sqrt{N} \Omega$, where $N$ is the total number of atoms in the ensemble, and $\Omega$ is the single-atom Rabi frequency~\cite{brion2007quantum, ebert2015coherence}. However, deterministic control over $N$ in experimental settings is challenging, making it difficult to implement a robust $\pi$ pulse between these two states. Additionally, thermal motion and the inter-atomic collisions within the ensemble induce rapid dephasing between $\ket{G}$ and $\ket{A}$ with characteristic timescales ranging from a few microseconds~\cite{Dudin2012strongly, Zeiher2015microscopic} to tens of microseconds ~\cite{Mei2022trapped}. As a result, despite long-standing interest in encoding quantum information within collective states of atomic ensembles~\cite{brion2007quantum,brion2008error,guo2020optimized}, achieving precise and high-fidelity manipulation between the $\ket{G}$ and $\ket{A}$ states remains a significant challenge. In SM, we further discuss the impact of residual $V_{Rr}$ on fidelities.

Our approach addresses these challenges by relying solely on high-fidelity initialization of $\ket{A}$ and precise quantum control over the two Rydberg states, $\ket{A}$ and $\ket{R}$. For state initialization, we use stimulated Raman adiabatic passage (STIRAP) to prepare the atomic ensemble in the $\ket{A}$ state. Compared to a direct $\pi-$pulse, STIRAP offers a robust and reliable process that is resilient to fluctuations in atom number, as well as instabilities in laser power and frequency~\cite{kuklinski1989adiabatic,petrosyan2013stimulated,xu2021fast}. After initialization, coherent control over $\ket{A}$ and $\ket{R}$ is achieved using transitions in the microwave domain, making the effects of thermal motion negligible. Although the Rabi coupling between $\ket{R}$ and $\ket{A}$ does not benefit from collective enhancement, the large transition dipole moments between nearby Rydberg states enable fast state manipulation on nanosecond timescales~\cite{xu2021fast, spong2021collectively}, with a dephasing time exceeding $\unit[15]{\mu s}$, which is sufficient for our proposed schemes.

Controlling interactions between atomic ensembles and individual atoms is also crucial. This can be achieved by exploiting the tunability of the strong Rydberg pair interactions between Yb and Rb atoms through the selection of appropriate Rydberg states, adjustments of external DC electric fields, and control over the array geometry. Here, we focus on the regime where $V_{Rr}\gg V_{Ar}, V_{rr}, V_{RR}, V_{AA}, V_{AR}$ at the nearest interatomic spacing, where $V_{xy}$ denotes the Rydberg interaction strength between Rydberg states $\ket{x}$ and $\ket{y}$. A promising approach to achieve this is by using interspecies F\"orster resonances, which brings a Rb-Yb Rydberg pair state into resonance with another Rydberg pair state via the DC Stark shift~\cite{beterov2015rydberg}. This results in a strong on-resonant dipole-dipole interaction, $V_{Rr}(d)~\sim 1/d^3$, while other interaction terms, such as $V_{Ar}$, $V_{rr}$ and $V_{AA}$, remain in the van der Waals (vdW) regime, where they decay more rapidly as $1/d^6$. Electrically tunable Rydberg interactions between dual species have already been demonstrated experimentally~\cite{ryabtsev2010observation, nipper2012highly, Ravets2014coherent, ravets2015measurement}, and a recent work~\cite{anand2024dualspecies} has used interspecies Rydberg interactions to generate quantum entanglement and transfer quantum states between species. While further experimental and theoretical investigations are needed to determine the optimal Rydberg state choices for Yb-Rb pair interactions, the distinct scaling behaviors of these interaction terms should allow the desired interaction hierarchy to be achieved. In sections~\ref{single-qubit gate},~\ref{multi-qubit gate}, and~\ref{qubit readout}, we discuss gate operations and measurement protocols under these conditions.  


\section{Reconfigurable local single-qubit gate operation}
\label{single-qubit gate}

Performing rapid and reconfigurable local gate operations on a subset of atoms is crucial for practical quantum computing but remains a significant challenge in neutral-atom-based quantum processing architectures. A common approach involves transporting atoms to a separate zone where global laser beams manipulate the target atoms~\cite{Bluvstein2022quantum, Bluvstein2024logical}. However, this method suffers from a transport time that is significantly longer than the gate operation time, and the total duration of transport may scale unfavorably with the array size. To overcome this limitation, several techniques have been developed to achieve rapid configurability without atom shuttling. These include applying local light shifts~\cite{chen2023continuous, Labuhn2014single, Burgers2022controlling}, selective shelving of atoms into other metastable states using site-selective light shifts~\cite{lis2023midcircuit}, and selectively driving qubit transition with tightly focused light beams via a two-photon process~\cite{Graham2022multi-qubit, Bluvstein2024logical}. While promising, these methods face challenges such as laborious calibrations and reliance on tightly focused beams, which are sensitive to variations in polarization, power and alignment. Furthermore, directly applying local addressing beams to data qubits introduces extra decoherence due to spontaneous emission and heating, complicating their practical implementation.

In this section, we propose several schemes for achieving reconfigurable local quantum control using dual-type atom arrays. The central idea is to selectively populate Rydberg excitations within target atomic ensembles. Once these excitations are prepared, the evolution of data qubits becomes conditional on the state of nearby atomic ensembles, even when global qubit manipulation beams are used. Compared to existing strategies, our approach reduces technical overhead and mitigates unwanted effects on data qubits by leveraging the simplicity of selective control over the atomic ensemble states. This results in a robust and efficient solution for implementing reconfigurable local quantum operations.

\subsection{Selective control over atomic ensemble states}
Fig.~\ref{architecture}(b) illustrates the concept of achieving selective control over atomic ensembles through a combination of local addressing beams and global driving laser beams. The local addressing beams induce light shifts on target atomic ensembles, changing the energy separation between $\ket{G}$ and $\ket{A}$. Starting with all atoms in their ground state, proper temporal control of $\Omega_p(t)$ and $\Omega_c(t)$, with the total frequencies of these beams resonant with the $\ket{G}$ $\leftrightarrow$ $\ket{A}$ transition, adiabatically evolves the state of atomic ensembles without addressing beams into the state $\ket{A}$, where a single Rydberg excitation is present within the ensemble. In contrast, the state of atomic ensembles exposed to addressing beams remains unchanged.

The fidelity of selective control over atomic ensembles depends on the magnitude of the light shift induced by the addressing beams. In previous work, off-resonant addressing beams were applied directly to data qubits, coupling one of the qubit states to a third state and inducing an AC Stark energy shift proportional to $\Omega_a^2/\Delta_a$, while also causing photon scattering at a rate proportional to $1/\Delta_a^2$. Here, $\Omega_a$ and $\Delta_a$ denote the Rabi rate and detuning of the addressing beams, respectively. To suppress heating and errors from light scattering while maintaining the desired energy shift, large detunings and high laser power are typically preferred, however, this introduces technical difficulties and limits scalability. In contrast, our proposed scheme does not rely on coherence between the $\ket{G}$ and the Rydberg state, allowing the use of near-resonant addressing beams that couples the ground state to the $\ket{5P}$ state. The strong dipole coupling of this transition introduces a large energy shift with a significantly lower laser power requirement. Importantly, throughout this process, the Yb data qubits remain in the ``dark'' and do not experience decoherence. A straightforward experimental approach is to use the optical tweezer beams themselves as addressing beams. Selective and reconfigurable control of these tweezer beams can be implemented using techniques demonstrated in~\cite{Zhang2024scaled}, where commercial optical modulators are used to selectively turn off subsets of optical tweezers on a timescale of approximately $\unit[20]{\mu s}$. By choosing a trapping light wavelength of $\unit[810]{nm}$, a beam waist $\unit[1.5]{\mu m}$, and a power of $\unit[4]{mW}$, 
we can initialize a subset of atomic ensembles towards $\ket{A}$ with fidelity of 99.2\%, which is limited by the finite lifetime of the Rydberg state, while keeping other atomic ensembles in their ground state with a fidelity of 99.8\%. 
Resetting the state of these atomic ensembles to $\ket{G}$ can be achieved by applying an optical pumping beam that couples the Rydberg states to a short-lived intermediate state or by turning on the optical tweezer traps, which repel the Rydberg excitation. This reset procedure allows for the subsequent round of gate operations on data qubits by selectively exciting a different set of atomic ensembles. Atom loss from the atomic ensembles is not a concern, as our protocols do not rely on the exact total number of atoms per ensemble.

\subsection{Ensemble-assisted local Z-gate}
\begin{figure}
    \centering
    \includegraphics[width=.9\linewidth]{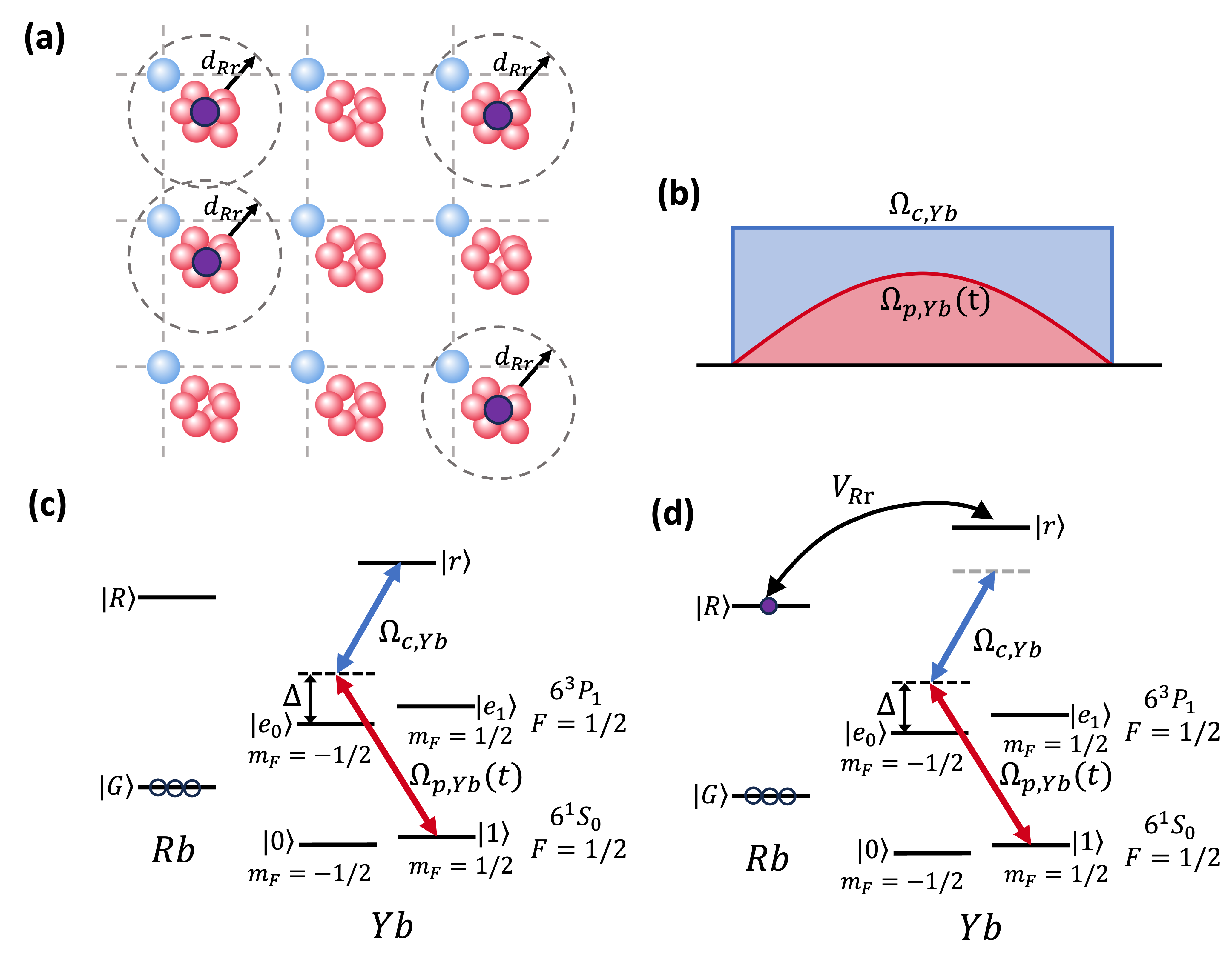}
    \caption{(a): Example configuration for implementing ensemble-assisted single-qubit operations. Individual addressability is achieved by selectively preparing a set of atomic ensembles adjacent to the target qubit in the state $\ket{R}$. The dashed circle indicates the Rydberg blockade radius $d_{Rr}$, within which the second Rydberg excitation is forbidden. (b)-(d): Temporal profile of global excitation beam pulses and the relevant energy levels for implementing a single-qubit $R_z(\theta)$ rotation on $^{171}$Yb atom. Two laser fields, $\Omega_{p,\text{Yb}}(t)$ and $\Omega_{c,\text{Yb}}$, couple the qubit state $\ket{1}$ to a Rydberg state $\ket{r}$ via an intermediate state. The pulse $\Omega_{p,\text{Yb}}(t)$ smoothly varies its intensity to satisfy the adiabatic condition, while $\Omega_{c,\text{Yb}}$ remains constant. The qubit evolution depends on the state of the nearby atomic ensemble. If the ensemble remains in the ground state $\ket{G}$, as shown in (c), adiabatic passage through the dark state ensures that the qubit returns to its initial state after the pulse sequence. In contrast, if the nearby ensemble is prepared in the excited state $\ket{R}$, as shown in (d), strong interactions between $\ket{R}$ and $\ket{r}$ break the dark-state condition, resulting a Pauli-Z rotation $R_z(\theta)$ of the qubit. The rotation angle $\theta$ is determined by the total area under $\Omega_{p,\text{Yb}}(t)$.}
    \label{fig:single-qubit}
\end{figure}

Selective control of atomic ensemble states enables conditional evolution of data qubits through interspecies interactions. Fig.~\ref{fig:single-qubit} illustrates the relevant energy diagrams and laser pulse sequence used to implement a rapid Pauli-Z rotation on data qubits with individual addressability. Two global driving beams are used: a $\sigma_-$ polarized light with a time-dependent Rabi frequency $\Omega_{p,\text{Yb}}(t)$ couples the qubit state $\ket{1}$ to an intermediate state $\ket{e_0}=\ket{6s6p^3P_1, F=1/2, m_F=-1/2}$ with detuning $\Delta$. A second laser with Rabi frequency $\Omega_{c,\text{Yb}}$ couples $\ket{e_0}$ to the Rydberg state $\ket{r}$, satisfying the two-photon resonance condition between $\ket{1}$ and $\ket{r}$. Due to selection rules, the qubit state $\ket{0}$ remains uncoupled.

As illustrated in Fig.~\ref{fig:single-qubit}(c) and (d), the evolution of data qubits under global driving depends on the state of the nearby atomic ensemble. When the atomic ensemble remains in the ground state $\ket{G}$, no interaction occurs between the data qubit and the ensemble. In this case, the data qubit state evolves under the Hamiltonian:

\begin{equation}
\begin{split}
     H =& -\hbar \Delta \ket{e_0}\bra{e_0} +\hbar\frac{\Omega_{p,\text{Yb}}(t)}{2}(\ket{1}\bra{e_0}+\ket{e_0}\bra{1})\\
     +&\hbar\frac{\Omega_{c,\text{Yb}}}{2}(\ket{e_0}\bra{r}+\ket{r}\bra{e_0}).
\end{split}
\end{equation}
For this Hamiltonian, there exists a dark state with eigenenergy $E=0$:
\begin{equation}
    \ket{d(t)} = \frac{\Omega_{c,\text{Yb}}}{\Omega_{\text{Yb}}(t)}\ket{1} - \frac{\Omega_{p,\text{Yb}}(t)}{\Omega_{\text{Yb}}(t)}\ket{r},
\end{equation}
where $\Omega_{\text{Yb}}(t)=\sqrt{\Omega_{c,\text{Yb}}^2+\Omega_{p, \text{Yb}}(t)^2}$. At $t=0$, when $\Omega_{p,\text{Yb}}=0$, this dark state is $\ket{1}$. In general, an arbitrary qubit state can be expressed as $\ket{\psi(t=0)}=\alpha\ket{0}+\beta\ket{1}=\alpha\ket{0}+\beta\ket{d(t=0)}$. By smoothly varying $\Omega_{p,\text{Yb}}(t)$, the qubit state adiabatically follows the evolution of the dark state, resulting in $\ket{\psi(t)}=\alpha\ket{0}+\beta\ket{d(t)}$. At the end of the pulse sequence, when $\Omega_{p,\text{Yb}}(t)$ is ramped back to $0$, the qubit state returns to its initial state without accumulating any additional phase, equivalent to an identity operation on the data qubits.

On the other hand, when the atomic ensemble is in state $\ket{R}$, strong Rydberg interaction between $\ket{R}$ and $\ket{r}$ induces a large energy shift, effectively decoupling $\ket{e_0}$ from $\ket{r}$. Under this condition, the Hamiltonian simplifies to:
\begin{equation}
    H=-\hbar\Delta\ket{e_0}\bra{e_0}+\hbar\frac{\Omega_{p,\text{Yb}}(t)}{2}(\ket{1}\bra{e_0}+\ket{e_0}\bra{1}).
\end{equation}
In a far-detuned dispersive regime ($\Delta\gg \Omega_{p,\text{Yb}}$), the state $\ket{1}$ acquires a phase dependent on the AC Stark shift $\Delta_{\mathrm{AC}}(t) = \Omega_{p,\text{Yb}}(t)^2/4\Delta$. By choosing the gate time $T$ such that $\int_0^T \Delta_{\mathrm{AC}}(t) dt=\theta$, a single-qubit phase rotation $R_Z(\theta)$ is implemented. 

\begin{table}[tb]
    \centering
    \caption{Infidelity analysis of individual addressing single-qubit gates. Here $\Omega_{\mathrm{max}}=2\pi\times \unit[15]{MHz}$, $\Omega_{c,\text{Yb}}=2\pi\times \unit[30]{MHz}$, $\Delta=2\pi\times \unit[90]{MHz}$, $V_{Rr}=2\pi\times \unit[200]{MHz}$ and the lifetimes of $\ket{e_0}$, $\ket{r}$, and $\ket{R}$ are $\unit[0.874]{\mu s}$, $\unit[100]{\mu s}$ and $\unit[200]{\mu s}$ respectively, which correspond to Rydberg states with principal quantum number $n_r\sim 70$ and $n_R\sim 80$.}
    Identity operation
    \begin{tabular}{|m{.7\linewidth} m{.3\linewidth}|}
    \hline
    Error source & Estimated error\\
    \hline \hline
    Non-adiabatic transition     &  0.148\% \\
    Data qubit Rydberg decay     &  0.090\% \\
    Intermediate state scattering  &  0.013\% \\
    \hline \hline
    Total fidelity & 99.75\%\\
    \hline
    \end{tabular}
    $R_z(\pi)$ rotation operation
    \begin{tabular}{|m{.7\linewidth} m{.3\linewidth}|}
    \hline
    Error source & Estimated error\\
    \hline \hline
    Intermediate state scattering     &  0.316\%\\
    Ancilla Rydberg decay     & 0.306\% \\
    Imperfect Rydberg blockade & $<$0.005\% \\
    Data qubit Rydberg decay  &  $<$0.001\%\\
    \hline \hline
    Total fidelity & 99.38\%\\
    \hline
    \end{tabular}
    \label{tab:single-qubit}
\end{table}

Thereby, the single-qubit Z-rotation gate is only applied to data qubits within the blockade radius of an ensemble prepared in the $\ket{R}$ state, realizing local quantum control using global driving beams. We numerically simulate the evolution of data qubits using a pulse shape $\Omega_{p,\text{Yb}}(t)=\Omega_{\text{max}}\sin(\pi t/T)$ with a gate time $T=8\pi \Delta/\Omega_{\mathrm{max}}^2$, which yields a $\pi$-rotation around the $z$-axis at $t=T$. 
A detailed analysis of infidelities due to various error sources is listed in Table~\ref{tab:single-qubit}. Our scheme achieves fidelities of 99.75\% for the identity operation and 99.38\% for the rotation gate, comparable to recent experimental results using shelving techniques~\cite{lis2023midcircuit}. Further improvements in fidelity could be achieved by implementing shortcut-to-adiabaticity methods~\cite{guery2019shortcuts} to reduce the total gate times (see SM for details).

\section{Ensemble-assisted Multi-Qubit Operation}
\label{multi-qubit gate}

Implementing native multi-qubit operations with high fidelity and efficiency significantly reduces the total number of gates required for complex quantum information processing protocols. This benefits practical applications in the NISQ era, advances the realization of fault-tolerant quantum computation~\cite{Molmer2011efficient, dlaska2022quantum, auger2017blueprint}, and enriches the toolbox for quantum simulation of Hamiltonians with many-body interactions~\cite{Weimer2010Rydberg}. While the schemes developed in~\cite{levine2019parallel} can be extended to the multi-qubit case, they require recalibration of beam parameters, as gates involving different number of qubits cannot be implemented in parallel with global operation beams. In this section, we present two schemes for performing multi-qubit gate operations within the dual-type atom array, which enable implementing different multi-qubit gates in parallel without requiring recalibrating.





\begin{figure}
    \centering
    \includegraphics[width=\linewidth]{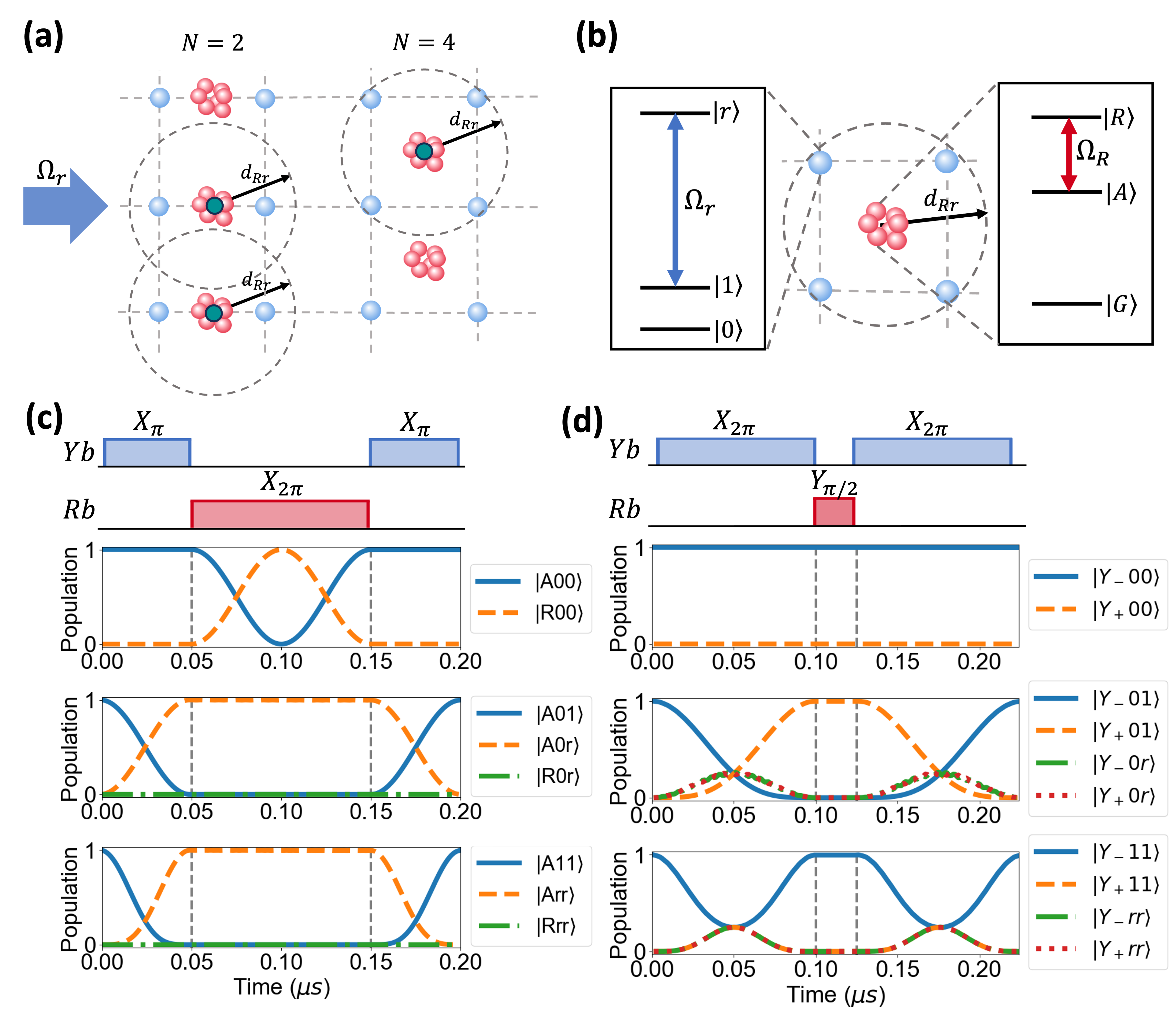}
    \caption{
    (a): Example configuration for implementing ensemble-assisted multi-qubit operations, where different configurations enable parallel execution of $C_{N-1} Z$ gate under global driving beams. Here, $\Omega_r$ denotes the effective Rabi coupling between the qubit state $\ket{1}$ and the Rydberg state $\ket{r}$. (b): Relevant energy level structure for multi-qubit operations. A global microwave field couples the two Rydberg states, $\ket{R}$ and $\ket{A}$, within the atomic ensemble. (c) and (d): Laser pulse sequence and numerically simulated population evolution for a controlled-Z ($CZ$) gate and a two-qubit phase-rotation gate ($R_{Z_1Z_2}(\pi/2)$), respectively. (c): For $CZ$ gate, when the initial state is $\ket{A00}$, the atomic ensemble undergoes a cyclic transition between $\ket{A}$ and $\ket{R}$, accumulating a $\pi$-phase shift. For initial states $\ket{A01}$, $\ket{A10}$ and $\ket{A11}$, the strong Rydberg blockade prevents the transition, leaving the ensemble staying in $\ket{A}$ and thus acquiring no phase shift. (d): For $R_{Z_1Z_2}(\pi/2)$, when the initial state is $\ket{Y_-00}$ or $\ket{Y_-11}$, where $\ket{Y_{-}}=(\ket{A}- i\ket{R})/\sqrt{2}$, the atomic ensemble remains in $\ket{Y_-}$ after the first $2\pi$ pulse, leading to a $\pi/4$ phase shift after the full sequence. For initial states $\ket{Y_-01}$ and $\ket{Y_-10}$, the atomic ensemble is transferred to $\ket{Y_+}$ by the first $2\pi$ pulse,  where $\ket{Y_+}=(\ket{A}+ i\ket{R})/\sqrt{2}$, accumulating a $-\pi/4$ phase shift after the full sequence. The numerical simulation uses the following parameters: $\Omega_r=\Omega_R=2\pi\times \unit[10]{MHz}$, Rydberg interaction strength $V_{Rr}=2\pi\times \unit[200]{MHz}$, and Rydberg state lifetimes of $\unit[100]{\mu s}$ for $\ket{r}$ and $\unit[200]{\mu s}$ for $\ket{R}$ and $\ket{A}$. }
    \label{fig:multi-qubit}
\end{figure}





\subsection{Multi-controlled-Z gate}
\label{CZ}
Here, we propose a scheme for implementing a multi-controlled-Z ($C_{N-1} Z$) gate, building on ideas from previous works~\cite{jaksch2000fast,isenhower2010demonstration,Isenhower2011multibit}. These earlier schemes require individually driving control and target qubits with sequential pulses, resulting in unequal times spent in the Rydberg states and increased dephasing, which limits gate fidelity. In contrast, our approach uses only global beams to drive the data qubits, ensuring that all data qubits spend an equal amount of time in their Rydberg states. This simplification enhances both experimental feasibility and gate fidelity. 

Fig.~\ref{fig:multi-qubit} illustrates the multi-qubit gate scheme. To perform a $C_{N-1} Z$ gate operation on target data qubits, $N$ data qubits are positioned within the blockade radius of an atomic ensemble. The protocol proceeds as follows: (i): the atomic ensemble is prepared in the $\ket{A}$ state. (ii): a global $\pi$-pulse is applied to all data qubits, coupling $\ket{1}$ to the Rydberg state $\ket{r}$. (iii): a global microwave $2\pi$-pulse is applied to the atomic ensemble, coupling the $\ket{A}$ and $\ket{R}$ states. Due to the Rydberg blockade effect, the atomic ensemble undergoes a cyclic transition and acquires a $\pi$-phase shift only if all surrounding data qubits within the blockade radius were initially in $\ket{0}$. (iv): Finally, a second global $\pi$-pulse on data qubits brings them back from $\ket{r}$ to $\ket{1}$. At the end of this sequence, the atomic ensemble remains in $\ket{A}$, and the data qubits undergo a unitary transformation given by:
\begin{equation}
U= Z_1Z_2\cdots Z_N (I^{\otimes N}-2\ket{0}^{\otimes N}\bra{0}^{\otimes N}).
\end{equation}
This operation is equivalent to a $C_{N-1}Z$ gate, up to single-qubit Pauli-Z rotations. The same single-qubit Pauli-Z rotations are also experienced by spectator qubits. Therefore, a global virtual Pauli-Z operation can offset these additional single-qubit rotations on all data qubits. 


Fig.~\ref{fig:multi-qubit}(c) shows the evolution of two data qubits within the blockade radius of an atomic ensemble under a $CZ$ gate. The corresponding error analysis, detailed in Table~\ref{tab:two-qubit}, indicates a gate fidelity of 99.80\%, which is comparable to the latest experimental results~\cite{Evered2023high-fidelity,Ma2023high-fidelity,Scholl2023erasure}. The primary source of infidelity is Rydberg-state decay, which can be reduced by a factor of two using a cryogenic platform~\cite{schymik2021single, cantat2020longlived, pichard2024rearrangement}. Additionally, recent protocols can convert Rydberg decay errors into quantum erasure errors, enhancing the performance of quantum error correction~\cite{Wu2022erasure,sahay2023high}. Another source of error arises from imperfect Rydberg blockade, which leads to leakage into doubly excited Rydberg states and unwanted phase accumulation. This leakage can be mitigated through pulse-shaping techniques (see SM for details).



\subsection{Multi-qubit phase-rotation gate}
\label{phase-rotation gate}
In addition to the multi-qubit controlled-Z gate discussed above, our architecture features a novel protocol for implementing a multi-qubit phase rotation gate with full control over the rotation angle. This protocol begins by initializing the atomic ensemble in the $\ket{A}$ state. A $\pi/2$-rotation around the $x$-axis, $R_x(\pi/2)$, is applied to the atomic ensemble, creating a superposition of two Rydberg states as $(\ket{A}-i\ket{R})/\sqrt{2}$. Next, a global $2\pi$ pulse, coupling the $\ket{1}$ and $\ket{r}$ states, is applied to all data qubits. As in $C_{N-1}Z$ gate case, the data qubits undergo a cyclic transition, acquiring a $\pi$-phase shift (equivalent to a Pauli-$Z$ operation) only if the atomic ensemble is in the $\ket{A}$ state. This results in an entangled state:
\begin{equation}
\label{eq:stabilizer}
\frac{1}{\sqrt{2}}(\ket{A}Z_1Z_2\cdots Z_N\ket{\psi_d}-i\ket{R}I_1I_2\cdots I_N\ket{\psi_d}),
\end{equation}
To disentangle the data qubits from atomic ensemble, a rotation along the $y$-axis, $R_y(\theta)$, is applied to the atomic ensemble, followed by another global $2\pi$-pulse on the data qubits. The resulting state is:
\begin{equation}
\frac{1}{\sqrt{2}}(\ket{A}-i\ket{R})(\cos\frac{\theta}{2}I_1I_2\cdots I_N + i \sin\frac{\theta}{2} Z_1Z_2\cdots Z_N)\ket{\psi_d}.
\end{equation}
At this stage, data qubits are disentangled from the atomic ensemble, and undergo the unitary transformation:
\begin{equation}
U = \cos\frac{\theta}{2}I_1I_2\cdots I_N + i \sin\frac{\theta}{2} Z_1Z_2\cdots Z_N = e^{i\frac{\theta}{2}Z_1Z_2\cdots Z_N}.
\end{equation}
This unitary operation corresponds to an N-qubit phase rotation gate $R_{Z_1Z_2\cdots Z_N}(\theta)$. For the two-qubit case, up to single-qubit rotations $R_{Z_1}(\theta)R_{Z_2}(\theta)$, the operation becomes:
\begin{equation}
    U\rightarrow e^{i\theta/2}\left(\begin{matrix}
        1&0&0&0\\
        0&1&0&0\\
        0&0&1&0\\
        0&0&0&e^{i2\theta}
    \end{matrix}\right)
\end{equation}
which is equivalent to a controlled-phase gate (CPhase). Spectator data qubits remain unaffected throughout the operation. Fig.~\ref{fig:multi-qubit}(d) shows the simulated evolution of two data qubits within the blockade radius of $\ket{R}$ under a phase rotation operation $R_{Z_1Z_2}(\pi/2)$, with detailed error analysis provided in Table~\ref{tab:two-qubit}. The primary sources of error, Rydberg state decay and imperfect blockade are similar to those in the $CZ$ gate. However, the $R_{Z_1Z_2}(\pi/2)$ achieves higher fidelity due to the shorter duration that data qubits spend in the Rydberg state (see SM for detailed infidelity analysis). 

\begin{table}[htbp]
    \centering
    \caption{Infidelity analysis of multi-qubit gate operations. Same parameters as in Fig.~\ref{fig:multi-qubit} are used.}
    $CZ$ gate
    \begin{tabular}{|m{.7\linewidth} m{.3\linewidth}|}
    \hline
    Error source & Estimated error\\
    \hline \hline
    Data qubit Rydberg decay     &  0.15\% \\
    Atomic ensemble Rydberg decay     &  0.03\% \\
    Imperfect Rydberg blockade  &  0.02\% \\
    \hline \hline
    Total fidelity & 99.80\%\\
    \hline
    \end{tabular}
    $R_{Z_1Z_2}(\pi/2)$ gate
    \begin{tabular}{|m{.7\linewidth} m{.3\linewidth}|}
    \hline
    Error source & Estimated error\\
    \hline \hline
    Data qubit Rydberg decay     &  0.050\% \\
    Atomic ensemble Rydberg decay     &  0.048\% \\
    Imperfect Rydberg blockade  &  0.002\% \\
    \hline \hline
    Total fidelity & 99.90\%\\
    \hline
    \end{tabular}
    \label{tab:two-qubit}
\end{table}

\subsection{Parallel multi-qubit gate operation}


A key advantage of our scheme is its direct extensibility to multi-qubit gates without requiring additional recalibration, as required by some previous methods~\cite{Jandura2022timeoptimal, Evered2023high-fidelity}. The same pulse sequences can be used to implement $C_{N-1} Z$ and $R_{Z_1\cdots Z_N}(\theta)$ gates for different values of $N$. This flexibility simplfies quantum circuit design and enables the parallel implementation of various multi-qubit gates using global driving beams.

\begin{figure}[htbp]
    \centering
    \includegraphics[width=\linewidth]{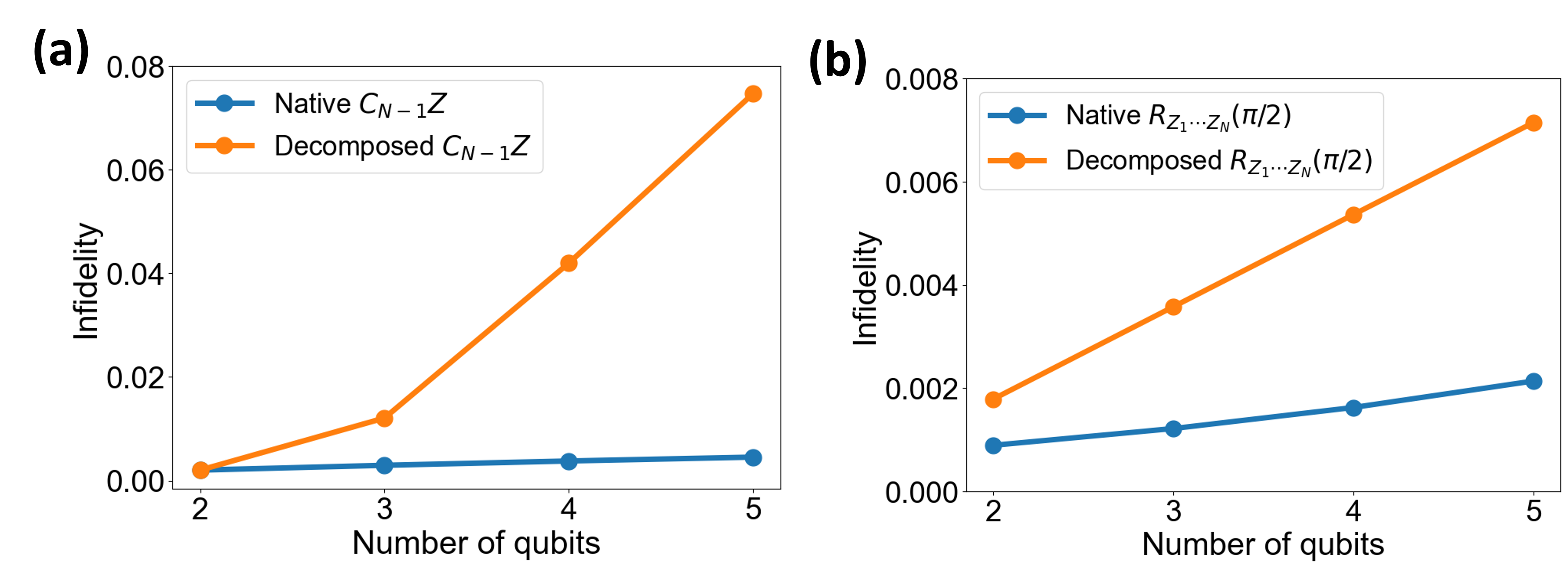}
    \caption{Infidelity comparison between the native implementation of multi-qubit gates and their decomposition into two- and single-qubit gates. (a) and (b) show the infidelities of $C_{N-1}Z$ gate and $R_{Z_1Z_2\cdots Z_N}$ gate for varying $N$, respectively. The decomposed quantum circuits are adapted from~\cite{barkoutsos2018quantum, balauca2022efficient}, with detailed discussion in SM.}
    \label{fig7}
\end{figure}

Compared to decomposing a multi-qubit gate into a series of single- and two-qubit gates, direct multi-qubit gate implementation offers significant advantages by reducing the number of required gate operations, leading to higher efficiency and lower error rates. To verify this, we examine the scaling of infidelity for the $C_{N-1}Z$ gate and $R_{Z_1\cdots Z_N}(\theta)$ gate as the number of qubits $N$ increases. We compare the infidelity of our proposed multi-qubit gate implementation with that of an equivalent gate decomposed into $CZ$ and single-qubit operations. As shown in Fig.~\ref{fig7}, the infidelity of our schemes scales approximately linearly with $N$, while the infidelity of the decomposed implementation grows exponentially, demonstrating the advantage of our approach (see more details in the SM).

It is also important to discuss the fault-tolerance of multi-qubit gates. The $C_{N-1} Z (N>2)$ and multi-qubit phase rotation gate belong to the non-Clifford group, which are costly to implement but essential for universal quantum computation. For two-dimensional stabilizer codes, it is well known that transversally implementable logical gates are restricted to the Clifford group \cite{bravyi2013classification}. In contrast, high-dimensional codes support fault-tolerant logical non-Clifford gates, including multi-controlled-Z gate \cite{vasmer2019three-dimensional, paetznick2013universal} and phase gate. Our proposed schemes, which use global qubit operation beams for versatile quantum control, combined with the geometrical reconfigurability of atom arrays, are well suited for demonstrating fault-tolerant non-Clifford gate operations. This opens up further opportunities to optimize the cost of logical quantum computation \cite{Wang2024efficient}.

\section{Rapid, non-demolition mid-circuit readout}
\label{qubit readout}
Many quantum technologies require the ability to perform fast, non-demolition mid-circuit readout, which occurs on timescales much shorter than the system's decoherence rate while affecting only a subset of the qubits without disturbing the rest. Examples of key applications of such techniques include repetitive quantum error correction~\cite{preskill1998reliable, steane1996multiple}, measurement-based quantum computation~\cite{briegel2009measurement, stephen2024universal}, and error-corrected quantum metrology~\cite{Kessler2014quantum, Dur2014improved, Zhou2020optimal} 

In neutral atom systems, conventional measurement methods involve state-selective atom loss followed by fluorescence imaging of the remaining atoms. High-fidelity imaging often takes tens of milliseconds in many machines, which is over 1,000 times slower than the gate speed. Furthermore, this destructive detection method prevents the reset and reuse of atoms, limiting the experimental repetition rates and making continuous operation modes challenging. Moreover, in single-species atom arrays, all atoms share the same resonant transition frequencies, resulting in scattered photons being reabsorbed by nearby atoms, which disturbs their quantum states. This significantly increases the difficulty in implementing advanced quantum protocols that require mid-circuit measurements on a subset of atoms. 

Significant efforts have been devoted to developing mid-circuit measurement techniques compatible with measurement-based quantum protocols and continuous operation~\cite{lis2023midcircuit, Norcia2023midcircuit, Huie2023repetitive}. Various approaches have been explored to achieve selective readout with low cross-talk, including using separate readout zones~\cite{Bluvstein2024logical}, shelving atoms in other states~\cite{Graham2023midcircuit, lis2023midcircuit}, and employing local light shifts~\cite{Norcia2023midcircuit}. However, optimizing for both fast readout and non-demolition detection remains challenging, as these requirements often conflict. Additionally, mitigating crosstalk during detection may require either long-distance atom shuttling or technically demanding shielding techniques. A dual-species atom array, when combined with strategies enabling controlled state transfer between species, naturally eliminates crosstalk. Since distinct atomic species have different optical transitions, measurements on one species can be performed without disturbing the other~\cite{singh2023midcircuit}. However, existing implementations still face limitations, particularly in achieving high-speed readout.

In this section, we show that dual-type dual-element arrays not only retain the advantage of crosstalk-free measurements, but also enable non-demolition detection, qubit reset, and reusability without compromising readout speed. By harnessing strong optical nonlinear effects in atomic ensembles and using them as detectors, we achieve high-fidelity single- and multi-qubit mid-circuit measurements within tens of microseconds.


\subsection{Rapid measurement on single-qubit states}
\label{single-qubit readout}

\begin{figure}
    \centering
    \includegraphics[width=\linewidth]{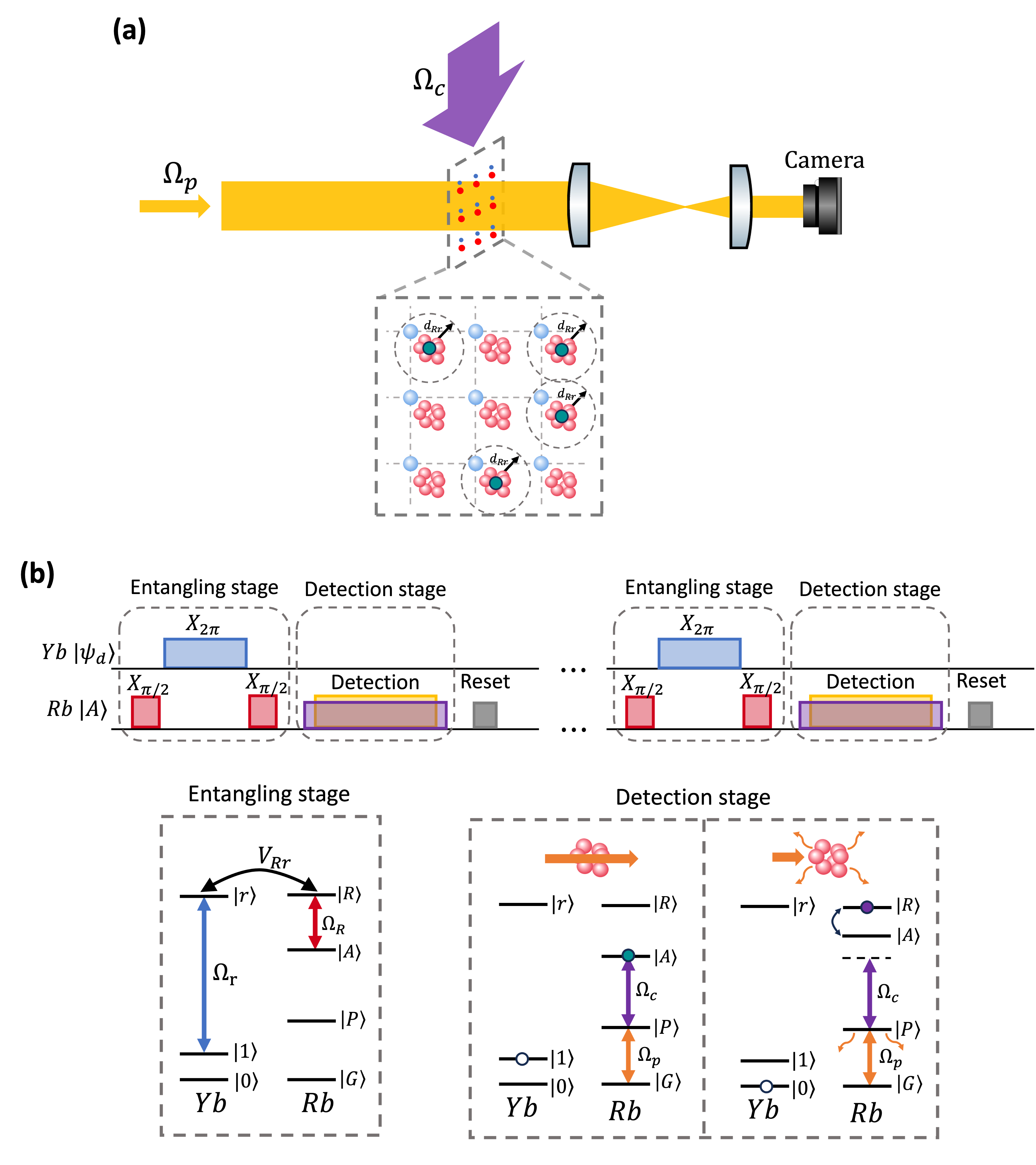}
    \caption{
    (a): Experimental setup for rapid, non-demolition mid-circuit readout of selected data qubits. Site-selective readout is achieved by selectively exciting the atomic ensemble adjacent to the target data qubits into its $\ket{A}$ state. A global probe beam illuminates the entire atom array, and its transmission is measured by using a camera. (b): Pulse sequence and relevant energy level structure for single-qubit readout. The readout protocol consists of two stages: an entangling stage and a detection stage. During the entangling stage, data qubits within the blockade radius of an atomic ensemble in its $\ket{A}$ state becomes entangled with Rydberg excitations within the ensemble. In the subsequent detection stage, an EIT-based scheme is used to rapidly distinguish different types of Rydberg excitations, as the interaction $V_{AR}$ within the atomic ensemble shifts the energy level, causing the fulfillment of the EIT condition to depend on the state of the atomic ensemble. Therefore, by monitoring the transmission of the probe beam on a camera, one can refer the target qubit states within a few microseconds. To further improve overall readout fidelity, the atomic ensemble can be reset to the ground state $\ket{G}$ after each cycle, allowing multiple entangling and detection rounds to enhance single-qubit readout fidelity.}
    \label{fig:repetitive_readout}
\end{figure}

We present a rapid, non-demolition mid-circuit single-qubit readout based on an electromagnetically-induced-transparency (EIT)-based scheme, originally introduced in~\cite{xu2021fast}, with several major improvements. Fig.~\ref{fig:repetitive_readout} illustrates the experimental setup. The atomic ensembles act as site-selective detectors by initializing them in the $\ket{A}$ state. The measurement scheme consists of two stages. In the first entangling stage, a pulse sequence, involving optical coupling between $\ket{1}$ and $\ket{r}$ and microwave coupling between $\ket{A}$ and $\ket{R}$ as shown in Fig.~\ref{fig:repetitive_readout}(b), entangles the data qubit state with the Rydberg excitation of the atomic ensemble: 
\begin{equation}
\begin{split}
-\ket{A}\frac{I-Z}{2}\ket{\psi_d}-i\ket{R}\frac{I+Z}{2}\ket{\psi_d}),
\end{split}
\end{equation}
where $(I+Z)/2=\ket{0}\bra{0}$ and $(I-Z)/2=\ket{1}\bra{1}$. Thus, measuring the Rydberg state of the atomic ensemble is equivalent to performing a Pauli-Z measurement on the data qubit. In the second detection stage, an EIT-based scheme ~\cite{xu2021fast} enables rapid differentiation between different Rydberg excitations within the ensemble on a microsecond timescale. Here, a probe beam $\Omega_p$, together with a control beam $\Omega_c$, couples $\ket{G}$ to $\ket{A}$. The fulfillment of the EIT condition depends on the state of the atomic ensemble state, resulting high probe beam transmission for state $\ket{A}$ and low transmission for state $\ket{R}$. However, in previous implementations, a dominant source of detection error was Rydberg state loss during this EIT-based detection, limiting fidelity to about 92\%. Our approach overcomes this limitation because, although Rydberg state decay within the atomic ensemble remains unavoidable, the overall measurement fidelity is not fundamentally constrained by this decay. In our measurement protocol, during the relatively long detection stage, data qubits remain in the ground-state manifold, and the measurement protocol itself is non-demolition. This allows the entire measurement protocol to be repeated multiple times to enhance fidelity. Assuming an initial single-shot fidelity of $F_d\approx0.92$, the overall readout fidelity can exceed $99\%$ after four repetitions (see SM for a detailed discussion on noise models).

\subsection{Multi-qubit joint measurement}

\label{stabilizer}



Beyond enabling rapid, non-demolition measurements of individual data qubits, our scheme can be extended to directly measure multi-qubit Pauli operators, enabling repetitive quantum error syndrome detection. When $N$ data qubits are placed within the blockade radius of the atomic ensemble and are subjected to the same pulse sequence as in Fig.~\ref{fig:repetitive_readout}, the system evolves to 

\begin{equation}
\begin{split}
&-\ket{A}\frac{I_1I_2\cdots I_N-Z_1Z_2\cdots Z_N}{2}\ket{\psi_d} \\
&-i\ket{R}\frac{I_1I_2\cdots I_N+Z_1Z_2\cdots Z_N}{2}\ket{\psi_d}
\end{split}
\end{equation}
Here, $(I_1I_2\cdots I_N+Z_1Z_2\cdots Z_N)/2$ and $(I_1I_2\cdots I_N-Z_1Z_2\cdots Z_N)/2$ project $\ket{\psi_d}$ onto the $+1$ and $-1$ eigenspaces of the multi-qubit Pauli operator $Z_1Z_2\cdots Z_N$, respectively. Thus, measuring the state of the atomic ensemble is equivalent to performing a measurement of the multi-qubit Pauli Z operator $Z_1Z_2\cdots Z_N$. Additionally, by applying single-qubit gate operations, other types of multi-qubit Pauli operators can also be measured. In section~\ref{QEC}, we discuss its application to quantum error syndrome detection in the surface code as a concrete example.




\section{Applications}


In this section, we discuss several concrete applications that leverage the unique features of the proposed dual-type, dual-element arrays.

\subsection{Quantum error correction}
\label{QEC}
Quantum information is highly susceptible to noise and hardware imperfections, making quantum error correction (QEC) crucial for achieving large-scale quantum computation~\cite{Dennis2002topological,fowler2012surface}. Fault-tolerant quantum computation encodes a logical qubit using multiple physical qubits. Errors on these physical qubits can be detected by measuring specific multi-qubit operators, known as stabilizers, without disturbing logical qubit states. Since decoherence during readout and decoding contributes significantly to logical errors, stabilizer measurement must be performed quickly to enable real-time quantum error correction~\cite{terhal2015quantum, Battistel2023realtime}. Furthermore, to mitigate measurement errors, multiple rounds of syndrome detection are typically required to ensure fault-tolerance~\cite{fowler2012surface, Horsman2012surface, Litinski2019gameofsurfacecodes}. Therefore, fast and repetitive stabilizer measurements, while preserving qubit coherence, are critical for scalable fault-tolerant quantum computation. 


\begin{figure}[htbp]
    \centering
    \includegraphics[width=\linewidth]{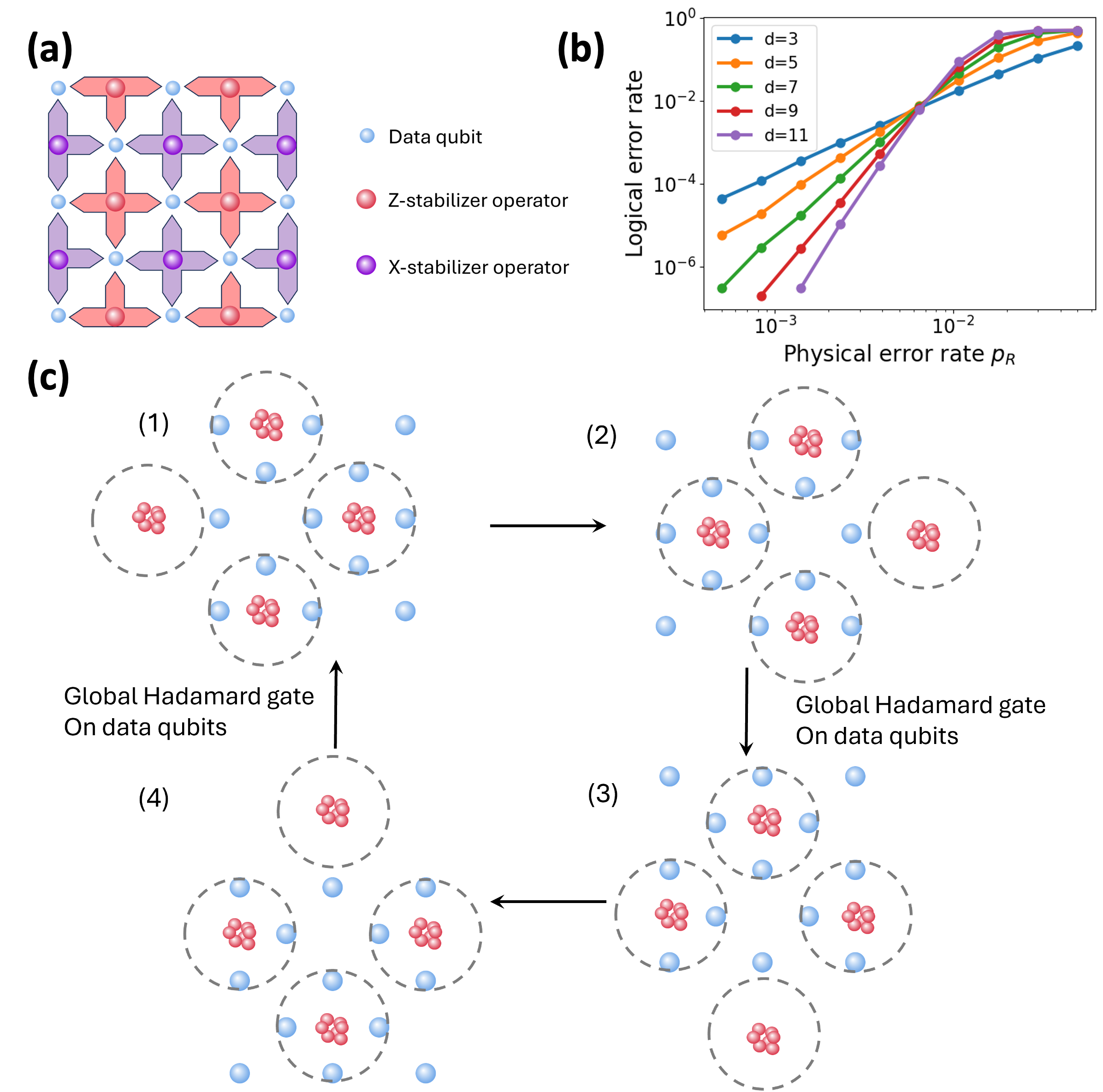}
    \caption{(a): Conceptual illustration of a distance-three surface code consisting of data qubits (blue circles), Z-type stabilizer operators (red circles) and X-type stabilizer operators (purple circles). In the experimental setup, stabilizer operator measurements are performed using atomic ensembles. (b) Logical qubit error rate as a function of physical error rate $p_R$ under the physical noise model, for various code distance. The error threshold is $p_{th}=0.7\%$. (c) Schematic of the implementation of quantum error syndrome detection for a distance-three surface code. The complete sequence of syndrome detection involves four steps. Step 1 and 2 measure Z-type stabilizers, while step 3 and 4 measure X-type stabilizers. A global Hadamard gate is applied to all data qubits between step 2 and 3 to modify the measurement basis. Between each step, atomic ensembles are transported short distances using a two-dimensional AOD. Dashed circles indicate the blockade radius of the atomic ensembles.}
    \label{fig:qec}
\end{figure}
Recently, stabilizer measurement using single ancillary qubits and multiple rounds of entangling gates has been demonstrated~\cite{Bluvstein2024logical}. However, this approach suffers from slow measurement time, with over 1000 times longer than gate operation time. Additionally, avoiding crosstalk between ancillary and data qubits requires long-distance atom shuttling, which imposes a significant time overhead and challenges the implementation of repetitive stabilizer measurements, crucial for efficient decoding~\cite{fowler2012surface}. In contrast, the multi-qubit operator measurement scheme described in Section~\ref{stabilizer}, enables efficient repetitive syndrome extraction within tens of microseconds, eliminating the need for long-distance atom shuttling.


To illustrate this, we discuss the stabilizer measurements in the surface code. In this scheme, stabilizers are divided into four groups and measured sequentially. Fig.~\ref{fig:qec}(c) depicts the experimental protocol for a distance-3 surface code. Step 1 and 2 measure Z-type stabilizers. Afterward, global Hadamard gates are applied on all data qubits, enabling steps 3 and 4 to measure the X-type stabilizers. This protocol extends to larger surface codes without additional time overhead, as such short-distance movements of atomic ensembles can be performed in parallel using a two-axis acousto-optic deflector (AOD). Importantly, since no quantum information is stored in the atomic ensembles during transport, they are resilient to atom loss and heating effects, enabling rapid and robust implementation.

To characterize the performance of this syndrome detection scheme, we consider a physical noise model that focuses on the finite lifetime of the Rydberg states in both data qubits and atomic ensembles during detection. A detailed analysis and discussion of this effective noise model can be found in SM. Using this effective noise model, we perform a Monte Carlo simulation to estimate the logical error rate for a standard surface code of distance $d$. This simulation samples Kraus operators at each error location and estimates the logical error rate after $d$ rounds of noisy syndrome measurements. To evaluate the tolerance of our scheme to the loss of Rydberg states and identify the error threshold, we vary the loss probability $p_R$ and $p_r$ together while keeping the single-shot EIT-based readout infidelity fixed at $p_{m,d}=0.08$ (initialization errors and single-qubit gate errors are not included). The resulting logical error rates for different code distances are shown in Fig.~\ref{fig:qec}(b). The error threshold, defined as the physical error rate below which the logical error rate decreases with increasing code distance, is estimated to be $p_{\mathrm{th}}\approx 0.7\%$, within reach of current experimental capabilities. Therefore, with stabilizer measurement significantly shorter than qubit coherence times, our platform is expected to demonstrate enhancement in logical qubit coherence through repetitive stabilizer measurements.

\subsection{Efficient long-range entanglement generation}


Long-range entangled states are fundamental in quantum physics, serving as the foundation for quantum error correction codes in quantum information processing~\cite{Dennis2002topological} and quantum-enhanced metrology~\cite{degen2017quantum, giovannetti2004quantum}. They also play an important role in topological order in condensed matter physics. However, generating such states in large-scale systems using unitary operations is challenging due to the finite Lieb-Robinson bounds~\cite{Lieb1972finite}, which limit the speed at which information propagates in quantum systems. This constraint makes it difficult to efficiently generate long-range entanglement on near-term quantum devices with limited coherence times. 

Incorporating mid-circuit measurements and feedforward into state preparation can overcome these limitations, allowing the generation of certain long-range entangled states in constant circuit depth~\cite{Lu2022measurement, Tantivasadakarn2024long-range, sahay2024finite,baumer2024efficient}. Due to the inherent randomness of quantum measurement outcomes, deterministic state preparation requires adaptive strategies, where each measurement result is processed in real-time to adjust subsequent operations. This feedforward process demands mid-circuit measurements to be significantly faster than the coherence times, along with reconfigurable local quantum control to dynamically modify the measurement basis and follow-up operations. Many theoretical proposals have demonstrated that this measurement-based approach can efficiently prepare a variety of entangled states, including GHZ state~\cite{verresen2021efficiently, zhu2023nishimori}, the Toric code~\cite{Raussendorf2005long-range,Aguado2008creation,Piroli2021quantum}, and more generally, all states that can be realized by CSS stabilizer codes~\cite{Bolt2016foliated}. 

Despite its theoretical appeal, experimental realization of these protocols on neutral-atom-based quantum architectures remains challenging. In particular, a key bottleneck is the difficulty of performing rapid mid-circuit measurements and reconfigurable local quantum control. In a recent paper \cite{evered2025probing}, mid-circuit measurement and feedforward were used to generate long-range entanglement. However, since the readout ancilla and data qubits were of the same atomic species, the ancilla must be physically transported to a spatially separated readout zone for local imaging. The slow transport process, combined with the loss of ancilla qubits during movement and readout, limited the overall entanglement generate rate and fidelity. 

Our platform offers distinct advantages for implementing these measurement-based protocols. By having rapid mid-circuit measurements of single- and multi-qubit operators, along with reconfigurable quantum control, we eliminate the need for long-distance transport of atoms and significantly reduce the readout time. Furthermore, any atom loss from the ensemble during readout does not compromise the entanglement generation rate. These capabilities open new avenues for exploring exotic topological orders and advanced quantum error correction codes. 


\begin{figure}
    \centering
    \includegraphics[width=.8\linewidth]{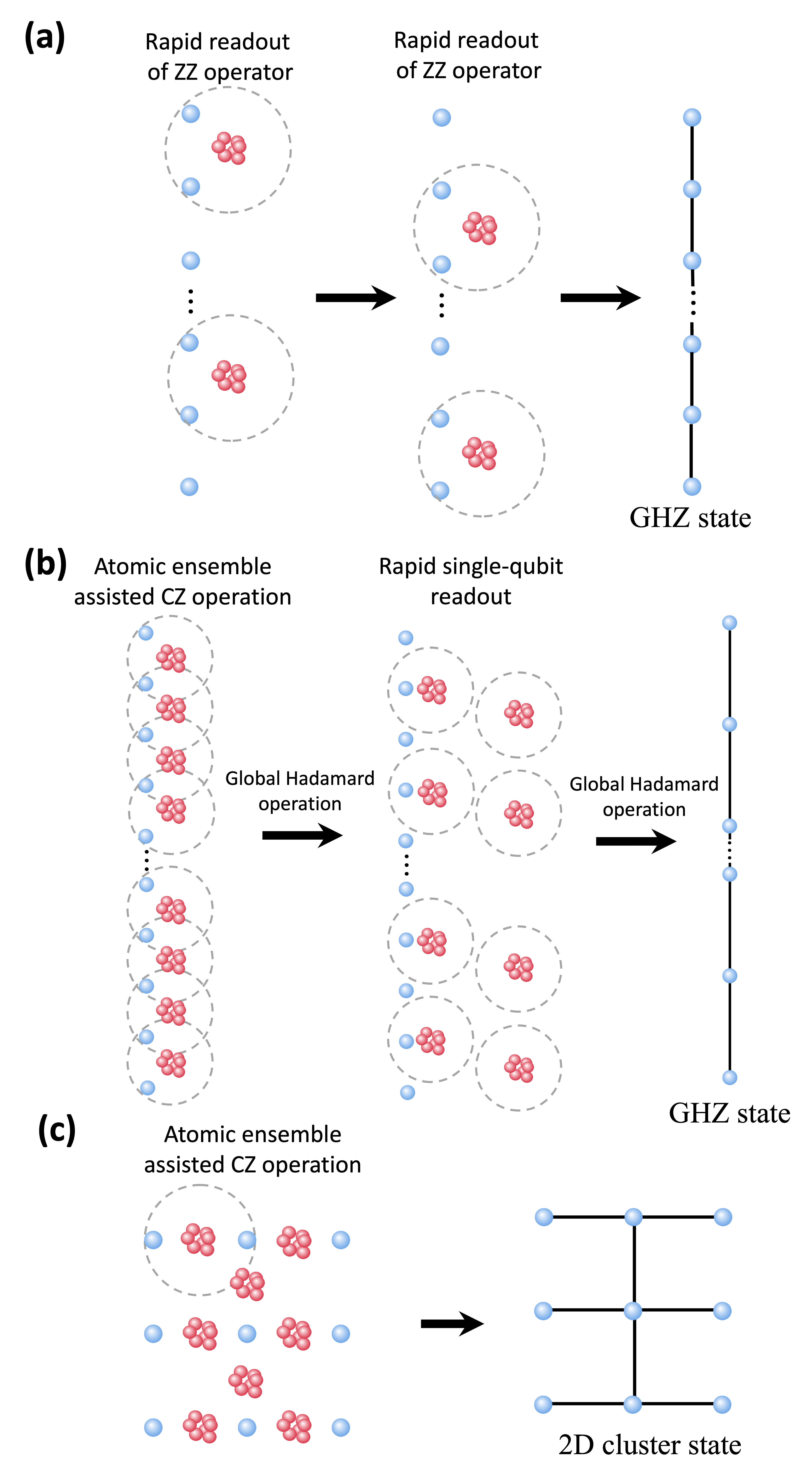}
    \caption{(a) and (b): Measurement-based protocols for efficient GHZ state preparation. The change in atomic ensembles' positions can be rapidly performed in parallel by using AODs. In (a), the GHZ state is generated through two rounds of parity measurements, followed by local correction operations if needed. (b) In an alternative approach, a cluster state is first prepared in a single step using ensemble-assisted CZ operations. Rapid single-qubit readout is then performed at odd sites, followed by local corrections to obtain the GHZ state. The global Hadamard operation is applied on all data qubits to rotate the measurement basis. (c) Configuration for 2D cluster state generation. A single round of ensemble-assisted CZ operations enables the preparation of the desired 2D cluster state. By using this cluster state as an input and applying measurement-based protocols, long-range entanglement in higher dimensions can be efficiently generated.}
    \label{GHZ}
\end{figure}


To elaborate on these capabilities, we consider the constant-depth preparation of an N-qubit GHZ state $\ket{\mathrm{GHZ}}=(\ket{0}^{\otimes N}+\ket{1}^{\otimes N})/\sqrt{2}$. GHZ states are valuable resources in quantum information processing, quantum communication, and quantum-enhanced metrology. Without the use of mid-circuit measurement, the depth of unitary quantum circuits required for GHZ state preparation scales linearly with the system size $N$. To avoid this, 
a measurement-based protocol as described in~\cite{verresen2021efficiently} can be used. This protocol begins by initializing all data qubits in $\ket{+}^{\otimes N}$, followed by pairwise two-body correlation operator measurements on $Z_1Z_2, Z_2Z_3\dots,Z_{N-1}Z_N$. These measurements can be implemented using the multi-qubit operator measurement scheme discussed in Section~\ref{phase-rotation gate}, where the correlation measurements are divided into two subsets and executed in two rounds (Fig.~\ref{GHZ}(a)). If all measurement outcomes are $+1$, the system is already in the GHZ state. Otherwise, a decoder determines the necessary correction operations, consisting of Pauli-X gates applied to a subset of data qubits. After these corrections, the final state is the desired GHZ state.

An alternative approach to preparing GHZ state efficiently is to use cluster states, which are symmetry protected topological phases, as the input state~\cite{Raussendorf2001one-way, Else2012symmetry}. In this case, the mid-circuit measurement requires only single-qubit readout. On our platform, after arranging the data qubits and ensembles into the desired configurations as shown in Fig.~\ref{GHZ}(b), a cluster state can be generated in a single step using the ensemble-assisted CZ gate, as described in Section~\ref{CZ}. The next step involves dynamically shifting the positions of atomic ensembles to enable rapid single-qubit readout at odd sites in their $X$-basis, using the scheme detailed in Section~\ref{single-qubit readout}. Conditioned on the measurement outcomes, the state of certain data qubits is flipped as necessary to produce the target GHZ state. 


While the above discussion focuses on one-dimensional GHZ state generation, measurement-based protocols can be extended to higher dimensions~\cite{verresen2022efficiently, Tantivasadakarn2024long-range}. The single-step preparation of two-dimensional cluster states, as illustrated in Fig.~\ref{GHZ}(c), combined with rapid single- and multi-qubit mid-circuit measurements, enables the experimental realization of a broad class of long-range entangled states in higher dimensions, including Toric code, fractons, and even certain types of non-Abelian topological orders. Our proposed architecture provides a scalable and flexible framework for generating and studying highly-entangled quantum states in programmable quantum systems, by using measurement as a resource.

\section{Conclusion and Outlook}
In this work, we proposed a novel quantum processing architecture based on dual-type dual-element atom arrays, designed to advance neutral-atom-based quantum technology. We developed protocols enabling reconfigurable individual quantum operations and rapid, non-demolition mid-circuit readout. A detailed performance analysis with realistic experimental parameters demonstrates gate fidelities of up to $99.5\%$ for individual addressed single-qubit gates and $99.9\%$ for both multi-qubit controlled-Z gate and phase rotation gates. Furthermore, the rapid, crosstalk-free mid-circuit readout methods support both single-qubit projective measurements and multi-qubit joint measurements within tens of microseconds, with expected fidelities exceeding $99\%$. Although our discussion primarily focuses on using Yb and Rb atoms, the proposed techniques are readily applicable to other dual-species atom arrays.

This platform unlocks a variety of novel directions for neutral-atom-array-based quantum computation and simulation. For example, there is growing interest in incorporating measurements into the study of quantum many-body dynamics, both in analog and digital settings. Distinct quantum phases, characterized by either volume-law or area-law entanglement entropy scaling depending on the measurements rate~\cite{Li2018quantum, Skinner2019measurement, Fisher2023random}, have been predicted. While this phenomenon has been extensively studied theoretically, experimental demonstrations remain limited~\cite{Noel2022measurement, Koh2023measurement, Google2023measurement}, especially in atom-array platforms. In particular, a major challenge in analog quantum simulation of spin models is that current detection methods are often much slower than the system's dynamical timescale, and atom shuttling for low-crosstalk mid-circuit readout is impractical, as changes in atom positions would also perturb the many-body Hamiltonian. Additionally, the properties of measurement-induced phases are inherently defined by quantum state trajectories conditioned on measurement outcomes. Consequently, experimentally probing such phenomena generally requires post-selection, which becomes infeasible for large-size many-body systems evolving over long timescales. This motivates the development of adaptive strategies that enable different trajectories to converge, eliminating the need for post-selection \cite{poyhonen2024scalable}. Our proposed architecture, with its capacity for fast, low-crosstalk measurements and reconfigurable individual addressability without shuttling data/spin qubits, overcomes key technical hurdles and provides an ideal platform for experimentally investigating measurement-induced quantum dynamics in quantum spin models.

Beyond quantum computation and simulation, an additional unique advantage of incorporating atomic ensembles is their ability to host Rydberg polaritons propagating inside, which can mediate robust exchange dynamics between atoms and photons~\cite{Thompson2017symmetry, khazali2019polariton}. This enables a cavity-free, deterministic atom-light quantum interface, allowing reliable quantum-state transfer between stationary atomic qubits and flying photonic modes. With this, we can realize programmable quantum interconnects, paving the way towards modular quantum computation~\cite{jiang2007distributed, monroe2014large}.






\section{Acknowledgments}
We thank Mark Saffman, Klaus M\o lmer, Anders S\o rensen and Shengpu Wang for insightful discussions. We also acknowledge Franklin J. Vivanco and Tao Alex Zheng for their valuable feedback on the manuscript. This work is supported by the Swiss National Science Foundation (SNSF) Starting Grant under grant number 218127.


\bibliography{reference}

\appendix

\end{document}


\begin{CJK*}{UTF8}{}
\title{SUPPLEMENTARY MATERIAL}
\maketitle
\end{CJK*}	

\section{Choice of Rydberg states and Yb-Rb F\"oster resonance}
\begin{figure}[htbp]
    \centering
    \includegraphics[width=\linewidth]{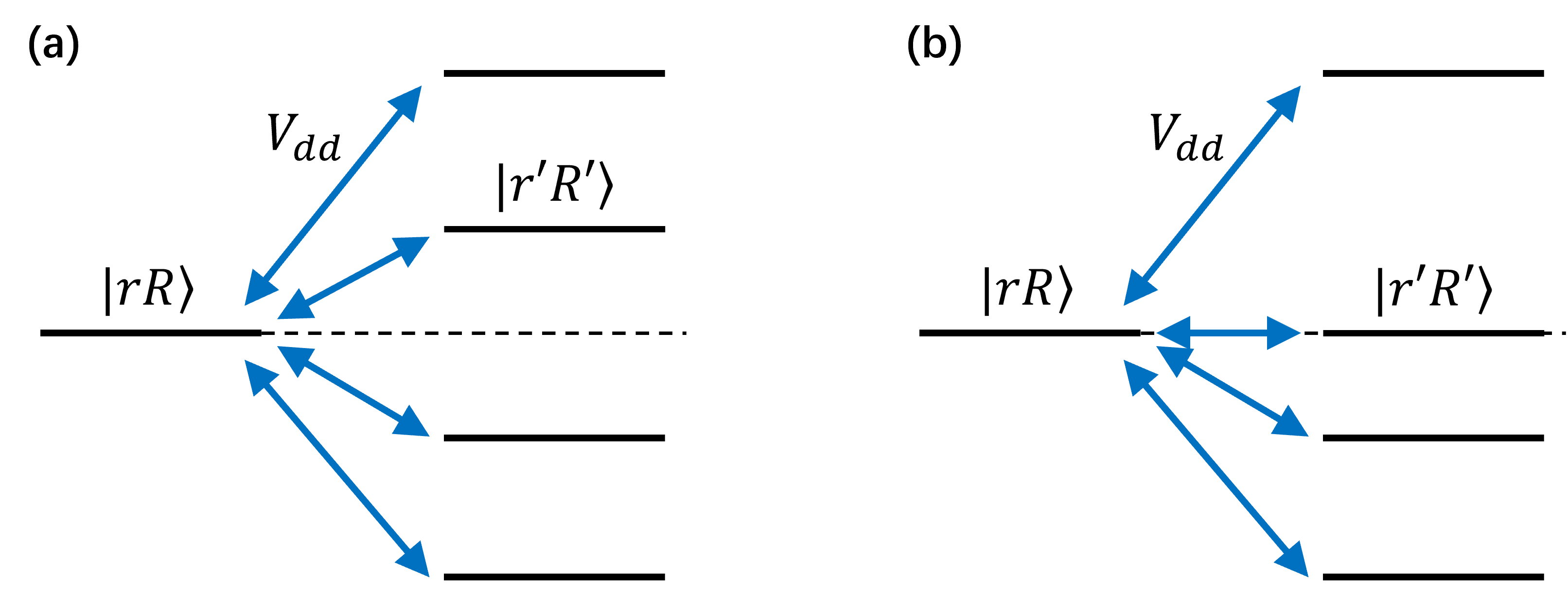}
    \caption{Two types of interaction between two Rydberg atoms. (a): Van der Waals interaction regime, where the interaction poential scales as $V(r) \propto 1/r^6$ and (b): F\"orster resonance regime, where $V(r) \propto 1/r^3$. Here, $r$ denotes the inter-atomic distance.} 
    \label{Forster_resonance}
\end{figure}
To experimentally implement the schemes proposed in this work with high-fidelity, proper choice on Rydberg states is essential. Three Rydberg states are involved, including one Yb Rydberg state (denoted as $\ket{r}$) and two Rb Rydberg states (denoted as $\ket{R}$ and $\ket{A}$). Interactions between them depend on the choices of Rydberg states, the interatomic distance, the magnitude of external electric fields, and the orientation of the interatomic axis with respect to the quantization axis. For the schemes discussed in this work, $V_{Rr}$ needs to be much larger compared to $V_{rr}$, $V_{RR}$, $V_{Ar}$, $V_{AA}$ and $V_{AR}$.


We propose to use F\"oster resonance to control the interaction between $\ket{r}$ and $\ket{R}$. For two Rydberg atoms, their dipole-dipole interaction is given by
\begin{equation}
    V_{dd} = \frac{1}{4\pi\epsilon_0}\left(\frac{\bm{d_1}\cdot\bm{d_2}-3(\bm{d_1}\cdot\bm{n})(\bm{d_2}\cdot\bm{n})}{\abs{\bm{r_1}-\bm{r_2}}^3}\right)
\end{equation}
where $\bm{d_{1,2}}$ are the electric dipole moments of the two atoms, $\bm{r_{1,2}}$ are their positions and $\bm{n}$ is the unit vector along $\bm{r_1}-\bm{r_2}$. 

If the energy difference between the pair state $\ket{rR}$ and any other pair states is much larger than the dipole-dipole interaction strength (Fig.~\ref{Forster_resonance}(a)), second-order perturbation theory results in a Van der Waals type interaction:
\begin{equation}
    H_{vdw} = \frac{C_6}{r^6}\ket{rR}\bra{rR}
\end{equation}
However, when the energy $\ket{rR}$ is degenerate with another pair state $\ket{r'R'}$ (\textit{i.e.,} $\Delta E=E_{rR}-E_{r'R'}=0$, as shown in Fig.~\ref{Forster_resonance}(b)), a strong dipole-dipole interaction arises:
\begin{equation}
    H_{FR} = \frac{C_3}{r^3}(\ket{rR}\bra{r'R'}+\ket{r'R'}\bra{rR})
\end{equation}
This resonance, known as F\"orster resonance, enables tunable interactions by adjusting external electric fields. A static electric field induces a DC Stark shift, modifying the value of $\Delta E$, and thereby allowing control over the interaction strength by shifting the pair states into or out of resonance. 

F\"orster resonance enhances the flexibility in selecting Rydberg state pairs for gate implementation and tuning their interaction strength. Due to its long-range nature, the F\"orster-type interaction $V_{Rr}$ decays more slowly than other pair-state interactions and dominates at relatively large interatomic distances compared to Van der Waals interactions. Moreover, this allows more data qubits to fit within the blockade radius of the atomic ensemble for multi-qubit control. Our schemes operate in the Rydberg blockade regime, that does not realy on the exact value of interaction strength. This relaxes the need for precise control over the external electric field and is less sensitive to the variations in interatomic distance, increasing both robustness and fidelity.


\section{Fidelity definition}
Session~\ref{single-qubit gate} and~\ref{Multi-qubit gate} define gate fidelity following ~\cite{Horodecki1999general} as:
\begin{equation}
    F = \int d\psi\ \mathrm{Tr}(U\ket{\psi}\bra{\psi}U^{\dagger}\rho),
\end{equation}
which averages over all pure input states. Here, $U$ is the target unitary operator, $\rho$ is the actual output given an input state $\ket{\psi}$, and the integral is taken over the uniform (Haar) measure $d\psi$ on the state space. 

While this fidelity definition is commonly used in analytical studies, directly computing the integral in numerical simulations is challenging. Therefore, when performing numerical gate fidelity analysis in Session III and IV of the main text, we use an equivalent definition adapted from~\cite{nielsen2002simple} as:
\begin{equation}
F = \frac{\sum_j\mathrm{Tr}(UU_j^{\dagger}U^{\dagger}\mathcal{E}(U_j))}{d^2(d+1)} + \frac{\mathrm{Tr}(UI_dU^{\dagger}\mathcal{E}(I_d))}{d(d+1)}
\end{equation}
where $U$ is the ideal unitary transformation matrix, $\mathcal{E}$ is the observed quantum operation, which can be computed numerical, and $d$ is the dimension of the system's Hilbert space. The identity operator of dimension $d$ is denoted as $I_d$, and $U_j$ represents a basis of unitary operators for a qudit. For a single-qubit gate $d=2$, we choose $U_j$ to be Pauli matrices $I,X,Y,Z$, and for two-qubit gate $d=4$, the $U_j$ is selected from the set $\{I,X,Y,Z\}^{\otimes 2}$.

\section{Ensemble-assisted local single-qubit gate}
\label{single-qubit gate}
In this section, we conduct a detailed fidelity analysis of the single-qubit gate scheme with individual addressibility, as presented in session III B of the main text. The laser pulse applied to data qubits follows the form $\Omega_{p,\text{Yb}}(t)=\Omega_{max}\sin(\pi t/T)$, where the gate time is  given by $T=8\pi\Delta/\Omega_{max}^2$. Ideally, this scheme results in an identical operation on the data qubits when the nearby atomic ensembles within their blockade radius remain in the ground state $\ket{G}$, while it induces a $\pi$-rotation when the nearby atomic ensembles are in the state $\ket{R}$. Below, we analyze the fidelity of these two scenarios separately.

\subsection{Identity operation fidelity}
In this case, two major sources of infidelity are the non-adiabatic transitions and the finite lifetime of Rydberg states.

\textbf{Non-adiabatic transitions:} For the identity operation, the data qubit state $\ket{1}$ adiabatically follows the dark state $\ket{d(t)}=\Omega_{c,\text{Yb}}/\Omega_{\text{Yb}}(t)\ket{1}-\Omega_{p,\text{Yb}}(t)/\Omega_{\text{Yb}}(t)\ket{r}$, where $\Omega_{\text{Yb}}(t)=\sqrt{\Omega_{p,\text{Yb}}(t)^2+\Omega_{c,\text{Yb}}^2}$. To analyze errors due to non-adiabatic transitions, we consider the Hamiltonian eigenstates outside the dark-state subspace, given by:
\begin{equation}
    \begin{split}
        \resizebox{\linewidth}{!}{$\ket{\psi_+}=\sqrt{\frac{\sqrt{\Delta^2+\Omega_{\text{Yb}}(t)^2}-\Delta}{2\sqrt{\Delta^2+\Omega_{\text{Yb}}(t)^2}}}\ket{e_0}+\sqrt{\frac{\sqrt{\Delta^2+\Omega_{\text{Yb}}(t)^2}+\Delta}{2\sqrt{\Delta^2+\Omega_{\text{Yb}}(t)^2}}}\ket{b}$} \\
        \resizebox{\linewidth}{!}{$\ket{\psi_-}=\sqrt{\frac{\sqrt{\Delta^2+\Omega_{\text{Yb}}(t)^2}+\Delta}{2\sqrt{\Delta^2+\Omega_{\text{Yb}}(t)^2}}}\ket{e_0}-\sqrt{\frac{\sqrt{\Delta^2+\Omega_{\text{Yb}}(t)^2}-\Delta}{2\sqrt{\Delta^2+\Omega_{\text{Yb}}(t)^2}}}\ket{b}$} \\
    \end{split}
\end{equation}
with corresponding eigen-energies
\begin{equation}
    E_{\pm}=\frac{-\Delta\pm\sqrt{\Delta^2+\Omega_{\text{Yb}}(t)^2}}{2}
\end{equation}
where $\ket{b}=\Omega_{p,\text{Yb}}(t)/\Omega_{\text{Yb}}(t)\ket{1}+\Omega_{c,\text{Yb}}/\Omega_{\text{Yb}}(t)\ket{r}$. Under the time-dependent Hamiltonian evolution, the dark state $\ket{d(t)}$ is coupled to $\ket{\psi_{\pm}}$ with coupling strength $C_{\pm}=\bra{\psi_{\pm}}\frac{d}{dt}\ket{d(t)}$. For $\Delta\gg\Omega_{\text{Yb}}(t)$, the population leakage towards non-dark states when the qubit is initially in $\ket{1}$ is given by
\begin{equation}
    P_{NA}\approx \frac{C_+^2}{E_+^2}+\frac{C_-^2}{E_-^2}\approx \frac{\Omega_{max}^6}{4\Omega_{c,\text{Yb}}^6}
\end{equation}
Averaging over all possible initial states, the averaged non-adiabatic error is $\epsilon_{NA}=\Omega_{max}^6/8\Omega_{c,\text{Yb}}^6$. To minimize this non-adiabatic error, the ratio $\Omega_{c,\text{Yb}}/\Omega_{max}$ should be kept as large as possible. For $\Omega_{c,\text{Yb}}/\Omega_{max}=2$ as used in our simulation, the estimated error due to non-adiabatic transition is around $1.95\times 10^{-3}$. 


Once a non-adiabatic transition occurs, population leakage into the short-lived intermediate state $\ket{e_0}$ introduces additional errors due to spontaneous decay. When the initial state is $\ket{1}$, the population in $\ket{e_0}$ during the gate operation is:
\begin{equation}
    P_e(t)=\left(\frac{C_+(t)^2}{E_+(t)^2}\abs{\braket{\psi_+(t)}{e_0}}^2+\frac{C_-(t)^2}{E_-(t)^2}\abs{\braket{\psi_-(t)}{e_0}}^2\right)
\end{equation}
The resulting average error due to intermediate-state decay is
\begin{equation}
    \epsilon_e=\frac{1}{2}\Gamma_e\int_0^TP_e(t)dt\approx\frac{\pi\Gamma_e\Omega_{max}^4(4\Omega_{c,\text{Yb}}^2+3\Omega_{max}^2)}{32\Delta\Omega_{c,\text{Yb}}^3(\Omega_{c,\text{Yb}}^2+\Omega_{max}^2)^{3/2}}
\end{equation} 
For sufficiently large values of $\Omega_{c,\text{Yb}}/\Omega_{max}$ and detuning $\Delta$, this effect remains minor. With $\Omega_{c,\text{Yb}}/\Omega_{max}=2$ as in our simulation, this error is approximately $6.6\times 10^{-3}\pi\Gamma_e/\Delta$.

\textbf{Rydberg state decay:} Another source of infidelity is the finite lifetime of the Rydberg state. This error can be estimated as $\Gamma_r\int_0^T P_r(t)dt$, where $P_r(t)$ is the population in the Rydberg state $\ket{r}$ at time $t$ and $\Gamma_r$ is the Rydberg state decay rate. When the initial state is $\ket{1}$, the state evolves as a dark state $\ket{d}$, leading to a Rydberg state population given by 
\begin{equation}
P_{r}(t)=\frac{\Omega_{p,\text{Yb}}(t)^2}{\Omega_{\text{Yb}}(t)^2}=\frac{\Omega_{p,\text{Yb}}(t)^2}{\Omega_{p,\text{Yb}}(t)^2+\Omega_{c,\text{Yb}}^2}
\end{equation}
Averaging over all input states, the resulting Rydberg decay error is
\begin{equation}
    \epsilon_r=\frac{\Gamma_r}{2}\int_0^TdtP_{r}(t)
=\frac{\Gamma_r T}{2}\left(1-\frac{\Omega_{c,\text{Yb}}}{\sqrt{\Omega_{max}^2+\Omega_{c,\text{Yb}}^2}}\right).
\end{equation}
This error can be suppressed by increasing the ratio $\Omega_{c,\text{Yb}}/\Omega_{max}$. For $\Omega_{c,\text{Yb}}/\Omega_{max}=2$ as in our simulation, the error due to data qubit Rydberg decay is around $0.05\Gamma_rT$. 

\subsection{Z-rotation gate fidelity}
When the atomic ensemble is in the $\ket{R}$ state, a Z-rotation gate is applied to the the data qubit within its blockade radius. In this case, two additional sources of infidelity arise: decay from the intermediate state of data qubits and decay of the atomic ensemble Rydberg state.

\textbf{Intermediate state decay:} In the Z-rotation gate, qubits are off-resonantly coupled to an intermediate state with a finite lifetime. Similar to the analysis of Rydberg state decay, the error caused by intermediate state decay can be estimated as $\epsilon_e=\Gamma_e \int_0^T P_e(t)dt$, where $P_e(t)$ is the population in the intermediate state $\ket{e_0}$ and $\Gamma_e$ is the decay rate. For $\Delta\gg\Omega_{p,\text{Yb}}(t)$, the intermediate state population is approximately
\begin{equation}
    P_e(t)\approx \frac{\Omega_{p,\text{Yb}}(t)^2}{4\Delta^2}P_1,
\end{equation}
where $P_1$ represents the initial population in $\ket{1}$. Averaging over all possible initial states, the intermediate state decay error is
\begin{equation}
    \epsilon_{e}=\frac{1}{2}\Gamma_e\int_0^T P_e(t)dt = \frac{\Gamma_e\Omega_{max}^2T}{16\Delta^2}=\frac{\pi\Gamma_e}{2\Delta}
\end{equation}

\textbf{Rydberg decay in atomic ensembles:} To perform a Z-rotation gate on data qubits, the nearby atomic ensemble is in its Rydberg state $\ket{R}$. Due to the finite Rydberg lifetime, the atomic ensemble may decay during the following gate operation. If a decay occurs, the data qubit undergoes adiabatic evolution until the gate operation is completed at time $T$. Since the gate fidelity is determined by the final state of the data qubit, the atomic ensemble state is traced out in the fidelity analysis, leading to an average error: 
\begin{equation}
\begin{split}
\epsilon_R =& \frac{1}{3}\int_0^T\Gamma_R dt(2-\frac{\Omega_{c,\text{Yb}}^2}{\Omega_{c,\text{Yb}}^2+\Omega_{p,\text{Yb}}(t)^2}\\
&-\frac{\Omega_{c,\text{Yb}}\cos\phi(t)}{\sqrt{\Omega_{c,\text{Yb}}^2+\Omega_{p,\text{Yb}}(t)^2}})\\
    =&\frac{1}{3}\Gamma_RT(2-\frac{\Omega_{c,\text{Yb}}}{\sqrt{\Omega_{c,\text{Yb}}^2+\Omega_{max}^2}})
\end{split}
\end{equation}
where $\phi(t)=\int_t^T\Omega_{\text{Yb}}(\tau)^2/2\Delta \,d\tau$. As the decay from $\ket{R}$ does not directly disrupt the data qubit's state, the resulting error is smaller than the raw decay probability. For $\Omega_{c,\text{Yb}}/\Omega_{max}=2$, the error caused due to Rydberg decay is approximately $0.37\,\Gamma_R T$.

\subsection{Fidelity improvement with shortcut-to-adiabaticity}
For the local single-qubit operation scheme in Section III B, the Z-rotation fidelity is primarily limited by the long gate duration required for adiabatic evolution. To enhance the performance of this scheme, the shortcut-to-adiabaticity (STA) technique can be applied to reduce the gate time~\cite{Mortensen2018fast, guery2019shortcuts}. Generally, given a Hamiltonian $H_0(t)$, non-adiabatic transitions among its instantaneous eigenstates can be suppressed by introducing a counter-diabatic term:
\begin{equation}
\label{counter-diabatic}
    H_{cd} = i\sum_n \ket{\partial_t \phi_n}\bra{\phi_n}-\braket{\phi_n}{\partial_t\phi_n}\ket{\phi_n}\bra{\phi_n}
\end{equation}
where $\ket{\phi_n(t)}$ are the instantaneous eigenstates of $H_0(t)$. With this additional term, the total physical Hamiltonian $H(t)=H_0(t)+H_{cd}(t)$ ensures that the system undergoes no non-adiabatic transitions, even for very short pulse durations. 

In the conditional local Z-rotation scheme, when the atomic ensemble is in state $\ket{G}$, the system follows adiabatic evolution under the Hamiltonian given in Eq.1 in main text. For large intermediate-state detuning $\Delta\gg \Omega_{p,\text{Yb}},\Omega_{c,\text{Yb}}$, the intermediate state can be adiabatically eliminated, yielding:
\begin{equation}
\begin{split}
H_0(t)\approx&\frac{\hbar\Omega_{p,\text{Yb}}(t)^2}{4\Delta}\ket{1}\bra{1}+\frac{\hbar\Omega_{c,\text{Yb}}^2}{4\Delta}\ket{r}\bra{r}\\
    &+\frac{\hbar\Omega_{p,\text{Yb}}(t)\Omega_{c,\text{Yb}}}{4\Delta}(\ket{1}\bra{r}+\ket{r}\bra{1})\\
    =&\frac{\hbar\Delta_{\text{eff}}}{2}(\ket{1}\bra{1}-\ket{r}\bra{r})-\frac{\hbar\Omega_{p,\text{Yb}}(t)^2}{8\Delta}\ket{0}\bra{0}\\
    &-\frac{\hbar\Omega_{c,\text{Yb}}^2}{8\Delta}\ket{0}\bra{0}+\frac{\hbar\Omega_{\text{eff}}}{2}(\ket{1}\bra{r}+\ket{r}\bra{1})
\end{split}
\end{equation}
where $\Delta_{\text{eff}}=\frac{\Omega_{p,\text{Yb}}^2-\Omega_{c,\text{Yb}}^2}{4\Delta}$ and $\Omega_{\text{eff}}=\frac{\Omega_{p,\text{Yb}}\Omega_{c,\text{Yb}}}{2\Delta}$. The counter-diabatic Hamiltonian, obtained from the Eq.~\ref{counter-diabatic}, is:
\begin{equation}
    H_{cd}(t)=i\frac{\hbar\Omega_a}{2}(\ket{1}\bra{r}-\ket{r}\bra{1})
\end{equation}
where $\Omega_a=\frac{2\Dot{\Omega}_{p,\text{Yb}}(t)\Omega_{c,\text{Yb}}}{\Omega_{p,\text{Yb}}(t)^2+\Omega_{c,\text{Yb}}^2}$. 


In a real experimental setting, the states $\ket{1}$ and $\ket{r}$ are not directly coupled. An experimental implementation of the Hamiltonian $H(t)=H_0(t)+H_{cd}(t)$ requires precise temporal control over the phase difference between the probe light $\Omega_{p,\text{Yb}}$ and control light $\Omega_{c,\text{Yb}}$, which can be technically challenging. To circumvent the need on relative phase control, a frame transformation can be applied:
\begin{equation}
    U=e^{-i\gamma(t)/2}\ket{1}\bra{1}+e^{i\gamma(t)/2}\ket{r}\bra{r}
\end{equation}
where $\gamma(t)=\arctan\left(\frac{\Omega_a}{\Omega_{\text{eff}}}\right)$. This transformation converts the time-dependent relative phase $\gamma(t)$ into a time-dependent detuning $\Dot{\gamma}(t)$, which can be easily implemented by adjusting the temporal profile of the probe and control light amplitudes. The resulting total Hamiltonian is:
\begin{equation}
\begin{split}
    \Tilde{H}(t)=&\frac{\hbar\Tilde{\Delta}_{\text{eff}}}{2}(\ket{1}\bra{1}-\ket{r}\bra{r})+\frac{\hbar\Tilde{\Omega}_{\text{eff}}}{2}(\ket{1}\bra{r}+\ket{r}\bra{1})\\
    &-\frac{\hbar(\Omega_{p,\text{Yb}}(t)^2+\Omega_{c,\text{Yb}}^2)}{8\Delta}\ket{0}\bra{0}
\end{split}
\end{equation}
where $\Tilde{\Delta}_{\text{eff}}=\Delta_{\text{eff}}+\Dot{\gamma}(t)$ and $\Tilde{\Omega}_{\text{eff}}=\sqrt{\Omega_{\text{eff}}^2+\Omega_a^2}$ To experimentally realize an evolution close to $\Tilde{H}(t)$, a good choice of the pulse profiles are:
\begin{equation}
    \begin{split}
    \Tilde{\Omega}_{p,\text{Yb}}(t)=\sqrt{2\Delta\left(\sqrt{\Tilde{\Delta}_{\text{eff}}^2+\Tilde{\Omega}_{\text{eff}}^2}+\Tilde{\Delta}_{\text{eff}}\right)}\\
    \Tilde{\Omega}_{c,\text{Yb}}(t)=\sqrt{2\Delta\left(\sqrt{\Tilde{\Delta}_{\text{eff}}^2+\Tilde{\Omega}_{\text{eff}}^2}-\Tilde{\Delta}_{\text{eff}}\right)}
    \end{split}
\end{equation}
With these pulses, the resulting Hamiltonian becomes:
\begin{equation}
\begin{split}
    \Tilde{H}'(t)=&\frac{\hbar\Tilde{\Delta}_{\text{eff}}}{2}(\ket{1}\bra{1}-\ket{r}\bra{r})+\frac{\hbar\Tilde{\Omega}_{\text{eff}}}{2}(\ket{1}\bra{r}+\ket{r}\bra{1})\\
    &-\frac{\hbar\sqrt{\Tilde{\Delta}_{\text{eff}}^2+\Tilde{\Omega}_{\text{eff}}^2}}{2}\ket{0}\bra{0}
\end{split}
\end{equation}
At the end of the pulse sequence, the population remains unchanged, but an additional phase $\theta_a$ accumulates between qubit state $\ket{0}$ and $\ket{1}$, given by:
\begin{equation}
\begin{split}
    \theta_a=&\int_0^Tdt\left(-\frac{\Omega_{p,\text{Yb}}(t)^2+\Omega_{c,\text{Yb}}^2}{8\Delta}+\frac{\sqrt{\Tilde{\Delta}_{\text{eff}}^2+\Tilde{\Omega}_{\text{eff}}^2}}{2}\right)\\
    &-\frac{\gamma(0)-\gamma(T)}{2}
\end{split}
\end{equation}
which corresponds to a Z-rotation $R_Z(\theta_a)$.




When the atomic ensemble is in state $\ket{R}$, the state $\ket{1}$ accumulates a phase depending on the AC Stark shift $\Delta_{AC}=\Tilde{\Omega}_{p,\text{Yb}}^2/4\Delta$, given by:
\begin{equation}
    \theta_Z=\int_0^T\Delta_{AC}(t)dt=\theta_a+\int_0^T dt\frac{\Omega_{p,\text{Yb}}(t)^2}{4\Delta}
\end{equation}
Since the additional rotation $R_Z(\theta_a)$ appears in both case, it can be compensated virtually.

\begin{figure}
    \centering
    \includegraphics[width=\linewidth]{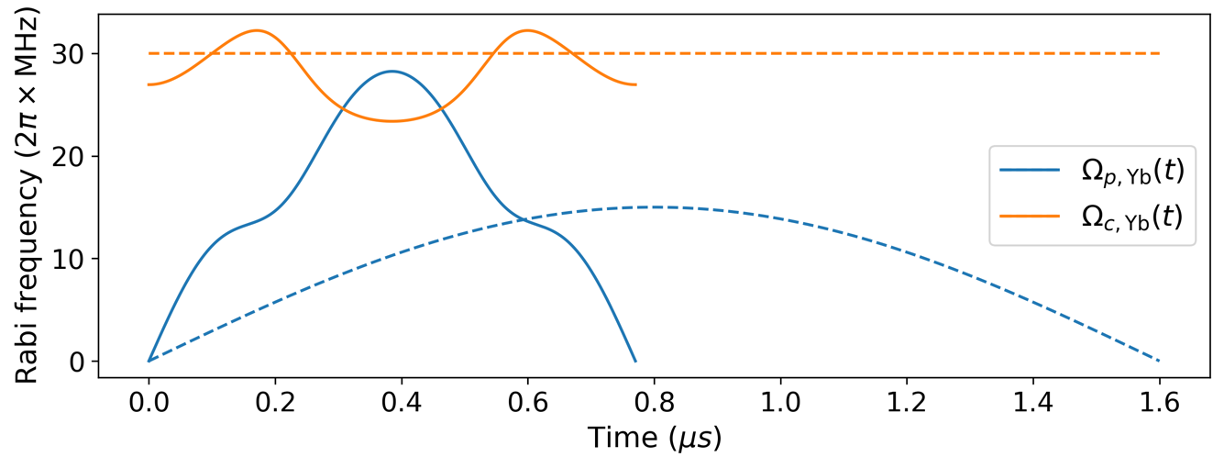}
    \caption{Pulse sequences for local single-qubit Z-rotation. The solid lines are the temporal profiles of $\Omega_{p,\text{Yb}}(t)$ and $\Omega_{c,\text{Yb}}(t)$ optimized using the shortcut-to-adiabaticity technique, while the dashed lines correspond to the pulse profiles used in the main text. By using STA, the gate time is reduced by a factor of two without compromising adiabaticity.}
    \label{fig:counter_diabatic}
\end{figure}

Using the same interaction strengths and Rydberg lifetimes as in Section A, this STA approach reduces the gate time from $1.6\mathrm{\mu s}$ to $0.77\mathrm{\mu s}$. Fig.~\ref{fig:counter_diabatic} shows the pulse shapes, which yield a $\pi$-phase rotation along the Z-axis. A detailed fidelity analysis is presented in Table~\ref{tab:counter_diabatic}. Compared to the scheme without the counter-diabatic term, the non-adiabatic transition error for identity operation is suppressed, and the $R_z(\pi)$ fidelity is improved due to the shorter gate duration. However, the data qubit Rydberg decay error for identity operations increases due to a larger Rydberg population caused by a higher $\Omega_{p,\text{Yb}}/\Omega_{c,\text{Yb}}$ ratio. Overall, the fidelity ($99.83\%$ for identity operation and $99.66\%$ for rotation operation) shows some improvement. Future work could explore optimal quantum control strategies that minimize Rydberg population time rather than only the total gate time.




\begin{table}[tb]
    \centering
    \caption{Error sources in individual addressing single-qubit gates using STA. The pulse sequences for $\Omega_{p,\text{Yb}}$ and $\Omega_{c,\text{Yb}}$ are shown in Fig.~\ref{fig:counter_diabatic}. We set the detuning to $\Delta=2\pi\times 130\mathrm{MHz}$.}
    Identity operation
    \begin{tabular}{|m{.7\linewidth} m{.3\linewidth}|}
    \hline
    Error source & Estimated error\\
    \hline \hline
    Non-adiabatic transition     &  0.001\% \\
    Data qubit Rydberg decay     &  0.122\% \\
    Intermediate state scattering  &  0.050\% \\
    \hline \hline
    Total fidelity & 99.83\%\\
    \hline
    \end{tabular}
    $R_z(\pi)$ rotation operation
    \begin{tabular}{|m{.7\linewidth} m{.3\linewidth}|}
    \hline
    Error source & Estimated error\\
    \hline \hline
    Intermediate state scattering     &  0.170\%\\
    Atomic ensemble Rydberg decay     & 0.168\% \\
    Non-perfect Rydberg blockade & $<$0.001\% \\
    Data qubit Rydberg decay  &  $<$0.001\%\\
    \hline \hline
    Total fidelity & 99.66\%\\
    \hline
    \end{tabular}
    \label{tab:counter_diabatic}
\end{table}

\subsection{Ensemble-assisted arbitrary single-qubit operation for hyperfine qubits} 

In session III of the main text, we discussed schemes for implementing local single-qubit Z-rotations for qubits encoded in nuclear spin degrees of freedom. These qubits typically have small energy splittings unless large magnetic fields are applied. In contrast, for alkali-metal atoms, qubits encoded in different hyperfine states have much larger energy separations, enabling independent coupling of $\ket{0}$ and $\ket{1}$ to the same intermediate state $\ket{e}$. This allows for an alternative scheme to implement local arbitrary single-qubit rotations, inspired by the work of M$\mathrm{\ddot{u}}$ller et al.~\cite{muller2009mesoscopic}.

\begin{figure}
    \centering
    \includegraphics[width=\linewidth]{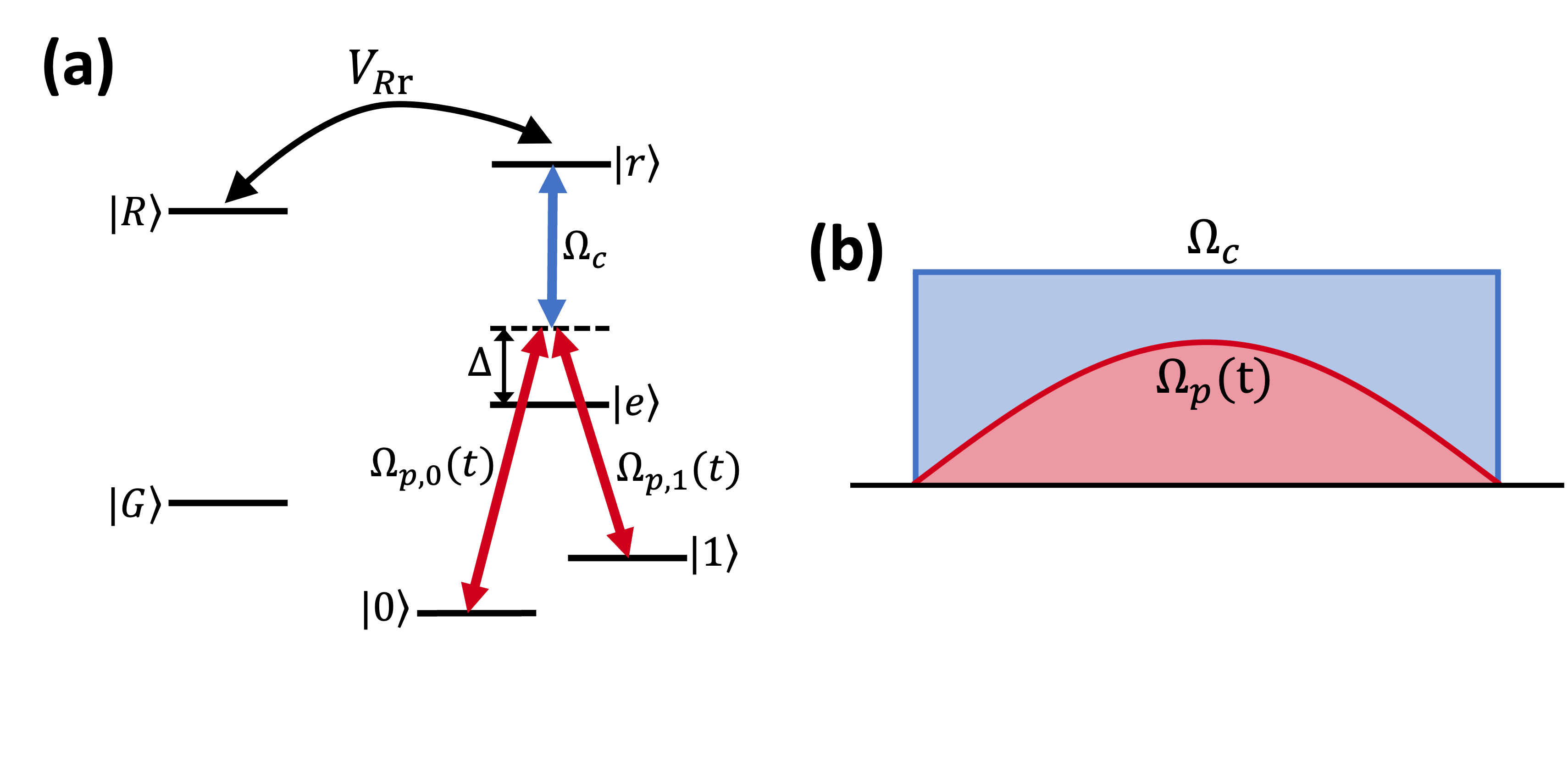}
    \caption{Ensemble-assisted arbitrary single-qubit rotation with individual addressability. (a) The relevant energy structure and laser pulse sequence for hyperfine qubits operations. Two laser beams, $\Omega_{p_0}(t)$ and $\Omega_{p_1}(t)$, couple the states $\ket{0}$ and $\ket{1}$ to an intermediate state $\ket{e}$ with detuning $\Delta$. Another laser beam $\Omega_c$ couples the state $\ket{e}$ to the Rydberg state $\ket{r}$ such that the two qubit states are in two-photon resonance with $\ket{r}$. The qubit state evolution depends on the state of the nearby atomic ensemble. When the atomic ensemble remains in the ground state $\ket{G}$, a smooth ramping on and off of $\Omega_{p_0}(t)$ and $\Omega_{p_1}(t)$ ensures that the data qubit adiabatically follows a dark state and returns to its initial state. On the other side, if the atomic ensemble is in state $\ket{R}$, the strong interaction between $\ket{R}$ and $\ket{r}$ destroys the dark state, causing the data qubit to undergo a Raman transition. (b) The temporal profiles of global driving beams.}
    \label{fig:single-qubit-arbitrary}
\end{figure}

Fig.~\ref{fig:single-qubit-arbitrary} shows the relevant energy levels and laser pulse sequences for the new ensemble-assisted single-qubit operation scheme. Two laser fields with Rabi frequencies $\Omega_{p_0}$ and $\Omega_{p_1}$ independently couple the qubit states $\ket{0}$ and $\ket{1}$ to the intermediate state $\ket{e}$ with a detuning $\Delta$ ($\Delta\gg \Omega_{p_0},\Omega_{p_1}$). A third laser couples $\ket{e}$ to a Rydberg state $\ket{r}$ with Rabi frequency $\Omega_c$. The qubit evolution depends on the state of the nearby atomic ensemble through $V_{rR}$. When the atomic ensemble is in ground state $\ket{G}$, it does not interact with the data qubit. The data qubit then evolves under the Hamiltonian:
\begin{equation}
\begin{split}
H=&-\hbar\Delta\ket{e}\bra{e}+\hbar\frac{\Omega_{p_1}(t)}{2}(e^{i\phi}\ket{1}\bra{e}+e^{-i\phi}\ket{e}\bra{1})\\
+&\hbar\frac{\Omega_{p_0}(t)}{2}(\ket{0}\bra{e}+\ket{e}\bra{0})+\hbar\frac{\Omega_c}{2}(\ket{e}\bra{r}+\ket{r}\bra{e})
\end{split}
\end{equation}
This Hamiltonian has two dark states with zero eigenenergy:
\begin{equation}
\label{dark}
\begin{split}
\ket{d_1(t)}=&\frac{\Omega_{p_1}(t)}{\Omega(t)}\ket{0}-e^{i\phi}\frac{\Omega_{p_0}(t)}{\Omega(t)}\ket{1}\\
\ket{d_2(t)}=&\frac{1}{\sqrt{1+x^2}}\left(\frac{\Omega_{p_0}(t)}{\Omega(t)}\ket{0}+e^{i\phi}\frac{\Omega_{p_1}(t)}{\Omega(t)}\ket{1}-x\ket{r}\right)
\end{split}
\end{equation}
where $x=\Omega(t)/\Omega_c(t)$ and $\Omega(t)=\sqrt{\Omega_{p_0}(t)^2+\Omega_{p_1}(t)^2}$. Initially, with $\Omega_{p_0}=\Omega_{p_1}=0$, the qubit state can be expressed as $\ket{\psi(t=0)}=\alpha\ket{d_1(t=0)}+\beta\ket{d_2(t=0)}$. By smoothly varying $\Omega_{p_0}$ and $\Omega_{p_1}$ while keeping their ratio $\Omega_{p_0}/\Omega_{p_1}$ and relative phase $\phi$ constant, the qubit state adiabatically follows the dark states as $\ket{\psi(t)}=\alpha\ket{d_1(t)}+\beta\ket{d_2(t)}$. At the end of the pulse sequence, when $\Omega_{p_0}$ and $\Omega_{p_1}$ are ramped back to zero, the qubit returns to its initial state without any additional phase accumulation, equivalent to an identity operation.

On the other side, when the atomic ensemble is in the Rydberg state $\ket{R}$, the strong Rydberg interaction $V_{Rr}$ shifts the energy of $\ket{r}$, suppressing the transition between $\ket{e}$ and $\ket{r}$. Under the condition $\Delta \gg \Omega_{p_0},\Omega_{p_1}$, the data qubit undergoes an effective two-photon Raman process, described by the Hamiltonian:
\begin{equation}
\begin{split}
H=&\frac{\hbar\Omega_{p_0}(t)^2}{4\Delta}\ket{0}\bra{0}+\frac{\hbar\Omega_{p_1}(t)^2}{4\Delta}\ket{1}\bra{1}\\
+&\frac{\hbar\Omega_{p_0}(t)\Omega_{p_1}(t)}{4\Delta}(e^{i\phi}\ket{1}\bra{0}+e^{-i\phi}\ket{0}\bra{1})
\end{split}
\end{equation}
By controlling $\Omega_{p_0}$, $\Omega_{p_1}$, and the relative phase $\phi$, arbitrary single-qubit operations can be applied to the data qubits. 


\section{Ensemble-assisted multi-qubit operation}
\label{Multi-qubit gate}
In this section, we present a detailed fidelity analysis of the multi-qubit operation discussed in session IV of the main text.

\subsection{Rydberg decay}
The error due to  Rydberg state decay is determined by the total time atoms spend in their Rydberg states. For data qubits, the Rydberg decay error is given by:
\begin{equation}
    \begin{split}
        \epsilon_r = \Gamma_r\frac{1}{2^n}\sum_{q\in \{0,1\}^n}\int_0^T P_{r,q}(t)dt
    \end{split}
\end{equation}
where $P_{r,q}(t)$ is population in the Rydberg state at time $t$ when the data qubits are initialized in state $\ket{q}$, and $\Gamma_r$ is the Rydberg decay rate for data qubits. For the CZ gate, the gate error is $\epsilon_r=\Gamma_r(\pi/\Omega_r+2\pi/\Omega_R)$, and for the two-qubit phase rotation gate $R_{Z_1Z_2}(\pi/2)$, the decay error is $\epsilon_r=\Gamma_r\pi/\Omega_r$, which has a higher fidelity compared to the CZ gate due to its shorter gate time. 

Both multi-qubit gate schemes rely on interactions between data qubits and atomic ensembles via their Rydberg states. Consequently, Rydberg decay in the atomic ensemble also impacts gate performance. While such decay does not directly destroy the quantum state of the data qubits, it alters the subsequent evolution. For the CZ gate, the gate duration is $T_R=2\pi/\Omega_r+2\pi/\Omega_R$, resulting an estimated decay error from the atomic ensemble as $\epsilon_R=\Gamma_R(3\pi/5\Omega_r+3\pi/5\Omega_R)$. For the $R_{Z_1Z_2}(\pi/2)$ gate, the gate duration is $T_R=4\pi/\Omega_r+\pi/2\Omega_R$, leading to an estimated error as $\epsilon_R=\Gamma_R(69\pi/40\Omega_r+\pi/5\Omega_R)$. In both cases, since only the final states of the data qubits is relevant to the gate fidelity, the error contribution from Rydberg decay in atomic ensemble Rydberg is smaller than the decay probability itself. 


\subsection{Non-perfect Rydberg blockade}


Multi-qubit gates are designed under the assumption of a perfect Rydberg blockade, where the Rydberg interaction is infinitely strong. However, in realistic experimental conditions, finite Rydberg interactions introduce two main effects: population leakage into doubly excited Rydberg states and dispersive phase shifts due to AC Stark effects. To access their impact on gate fidelity, we consider the case consisting of a single data qubit and an atomic ensemble, initially prepared in the state $\ket{1R}$. 

When Rydberg coupling is applied to the data qubits and a F\"{o}rster resonance between $\ket{r}$ and $\ket{R}$, the system evolves under the Hamiltonian:
\begin{equation}
\begin{split}
    H=&\frac{\hbar\Omega_r}{2}(\ket{1R}\bra{rR}+\ket{rR}\bra{1R})\\
    +&\hbar V_{Rr}(\ket{rR}\bra{r'R'}+\ket{r'R'}\bra{rR})
\end{split}
\end{equation}
A new dressed-state basis for $\ket{1R}$ and $\ket{r'R'}$ can be defined as 
\begin{equation}
\begin{split}
    \ket{\widetilde{1R}}&=\frac{2V_{Rr}}{\sqrt{\Omega_r^2+4V_{Rr}^2}}\ket{1R}-\frac{\Omega_r}{\sqrt{\Omega_r^2+4V_{Rr}^2}}\ket{r'R'}\\
    \ket{\widetilde{r'R'}}&=\frac{\Omega_r}{\sqrt{\Omega_r^2+4V_{Rr}^2}}\ket{1R}+\frac{2V_{Rr}}{\sqrt{\Omega_r^2+4V_{Rr}^2}}\ket{r'R'}.
\end{split}
\end{equation}
In this basis, the Hamiltonian becomes
\begin{equation}
    H=\frac{\hbar\sqrt{\Omega_r^2+4V_{Rr}^2}}{2}\left(\ket{\widetilde{r'R'}}\bra{rR}+\ket{rR}\bra{\widetilde{r'R'}}\right).
\end{equation}
Here, the dressed state $\ket{\widetilde{1R}}$ is a dark state that remains decoupled from $\ket{rR}$ and $\ket{\widetilde{r'R'}}$. The initial state $\ket{1R}$ can be expressed as:

\begin{equation}
    \ket{1R} = \frac{2V_{Rr}}{\sqrt{\Omega_r^2+4V_{Rr}^2}}\ket{\widetilde{1R}}+\frac{\Omega_r}{\sqrt{\Omega_r^2+4V_{Rr}^2}}\ket{\widetilde{r'R'}}.
\end{equation}
This indicates that population leakage into $\ket{rR}$ and $\ket{r'R'}$ may occur with probability up to $\Omega_r^2/(\Omega_r^2+4V_{Rr}^2)$ during gate operations. 

To avoid population to be remained at these doubly excited Rydberg states at the end of gate operations, the Rabi frequency $\Omega_r$ should be varied slower than $1/V_{Rr}$, ensuring adiabatic following of the dark state $\ket{\widetilde{1R}}$ throughout the pulse sequence. Moreover, since $\ket{\widetilde{1R}}$ is an eigenstate with zero eigenenergy, the dispersive shift from AC Stark effects, caused by imperfect Rydberg blockade, is zero. This dark-state protection strategy extends to systems with N data qubits and an atomic ensemble~\cite{Khazali2020fast}. Therefore, by utilizing F\"oster resonance and pulse shaping, the proposed multi-qubit gate scheme remains robust against Rydberg blockade imperfections and unwanted phase accumulation from AC Stark shifts.



\subsection{Rydberg interaction between $\ket{A}$ and $\ket{r}$}
So far, we have assumed no interaction between $\ket{A}$ and $\ket{r}$. However, because both are Rydberg states, their pair-state energy $\ket{rA}$ can shift due to the van der Waals interaction $V_{Ar}$. In principle, if $V_{Ar}$ is well-calibrated, this shift can be compensated by adjusting the laser frequency to maintain resonant coupling between $\ket{1A}$ and $\ket{rA}$. However, in our case, $\ket{A}$ is a collective excited state, where a single Rydberg excitation is delocalized across the atomic ensemble. This leads to fluctuations in $V_{Ar}$ due to the position uncertainty $\delta r$. Given that $V_{Ar}\propto r^{-6}$, the resulting variance in $V_{Ar}$ is $\delta V_{Ar}=6V_{Ar}\delta r/r_0$. For $\delta V_{Ar}\ll \Omega_r$, the associated gate errors can be estimated as: $\epsilon_{Ar}=2/5(\delta V_{Ar})^2(\pi/\Omega_r+2\pi/\Omega_R)^2=72/5V_{Ar}^2(\delta r/r_0)^2(\pi/\Omega_r+2\pi/\Omega_R)^2$ for the CZ gate, and $\epsilon_{Ar}=4/5(\delta V_{Ar})^2(\pi/\Omega_r)^2=144/5V_{Ar}^2(\delta r/r_0)^2(\pi/\Omega_r)^2$ for the phase rotation gate $R_{Z_1Z_2}(\pi/2)$. To achieve $\epsilon_{Ar}<0.001$, we require $\delta V_{Ar}/\Omega_r < 0.005$ for a CZ gate, and $\delta V_{Ar}/\Omega_r<0.011$ for a $R_{Z_1Z_2}(\pi/2)$ gate (assuming $\Omega_R=\Omega_r$). For an atomic ensemble with a diameter of $1\mu$m, inter-species spacing of $10\mu$m, and $\Omega_{r}=2\pi \times 20$MHz, we require $V_{Ar}$ to be smaller than $2\pi\times 0.33$MHz and $2\pi \times 0.73$MHz for the CZ gate and for the $R_{Z_1Z_2}$ gate respectively. Thanks to the fast decay of van der Waals interactions with distance, these constraints are experimentally achievable. Moreover, by harnessing the anisotropy of Rydberg pair interactions and optimizing the geometrical configurations, errors due to finite $V_{Ar}$ can be further suppressed. 


\subsection{Numerical simulation of multi-qubit gate fidelity}
\begin{figure}
    \centering
    \includegraphics[width=\linewidth]{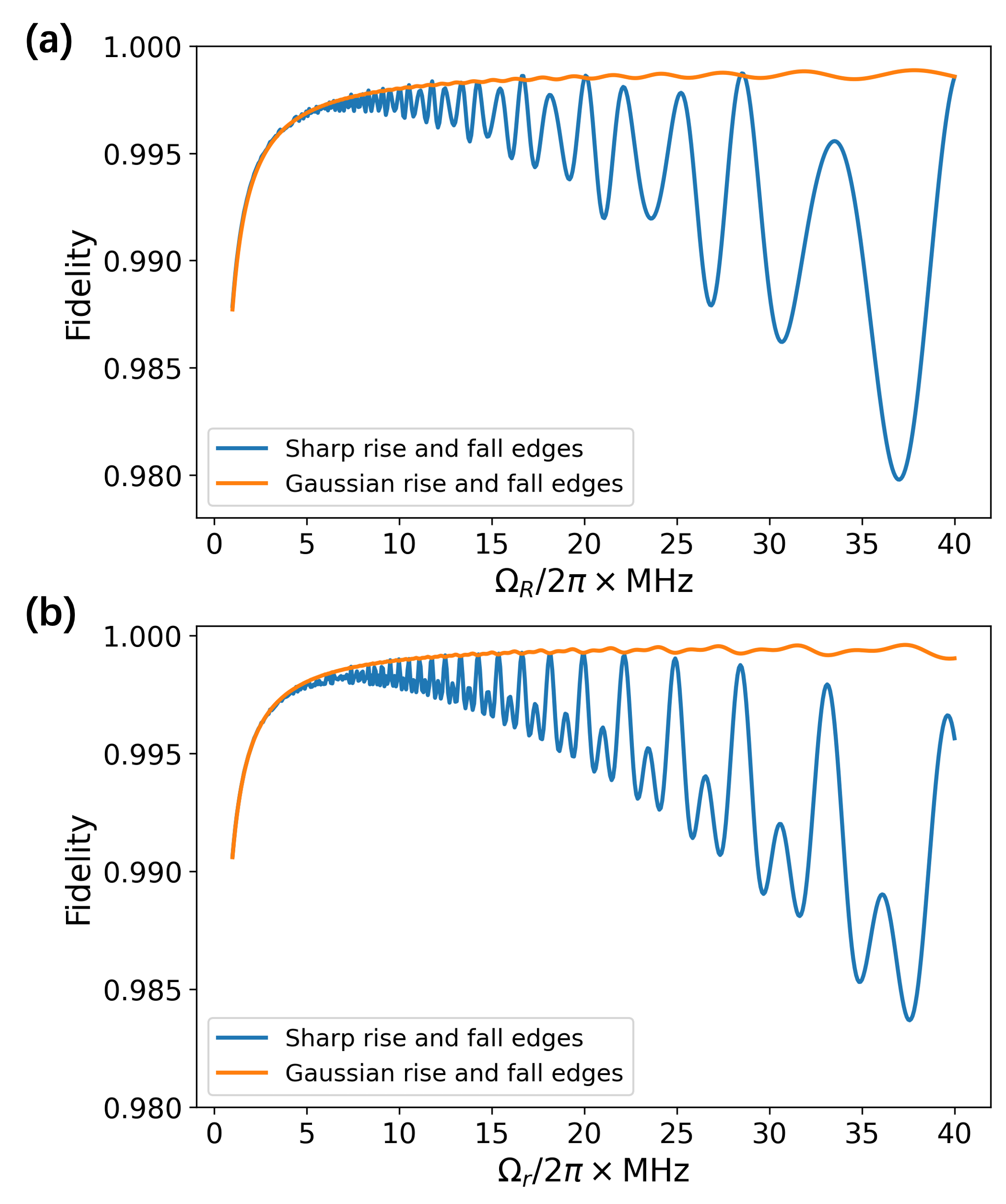}
    \caption{Simulated gate fidelity of (a) the controlled-Z gate $CZ$ and (b) the two-qubit phase rotation gate $R_{Z_1Z_2}(\theta)$ as a function of the Rabi frequency of the driving fields. At low Rabi frequencies, fidelity improves as the Rabi frequency increases. However, when using a drive pulse with sharp rise and fall edges (blue curves), fidelity starts to fluctuate due to insufficient Rydberg blockade and population leakage. In contrast, using a pulse with a smooth Gaussian rise and fall suppresses these oscillations, leading to more stable and high-fidelity gate operations (orange curves). }
    \label{fig:fidelity_multiqubit}
\end{figure}

Fig.~\ref{fig:fidelity_multiqubit} plots the simulated fidelity of the $CZ$ gate and the two-qubit phase rotation gate $R_{Z_1Z_2}(\theta)$, including the error sources discussed above. At low Rabi frequencies, the dominant source of error is Rydberg state decay due to the long duration of gate operations. As the Rabi frequency increases, the gate fidelity improves due to the shorter operation time. However, beyond a certain threshold, fidelity begins to fluctuate and decrease due to imperfect Rydberg blockade and population leakage into $\ket{r'R'}$. To address this issue, using a driving pulse with a smooth rise and fall, rather than a square pulse with sharp edges, significantly suppresses leakage into doubly excited Rydberg states and eliminates AC Stark shift induced phase. This approach enables robust, high-fidelity gate operations even at large Rabi frequencies, mitigating the impact of blockade imperfections.



\section{Decomposition of multi-qubit gate}

\begin{figure}[htbp]
    \centering
    \includegraphics[width=\linewidth]{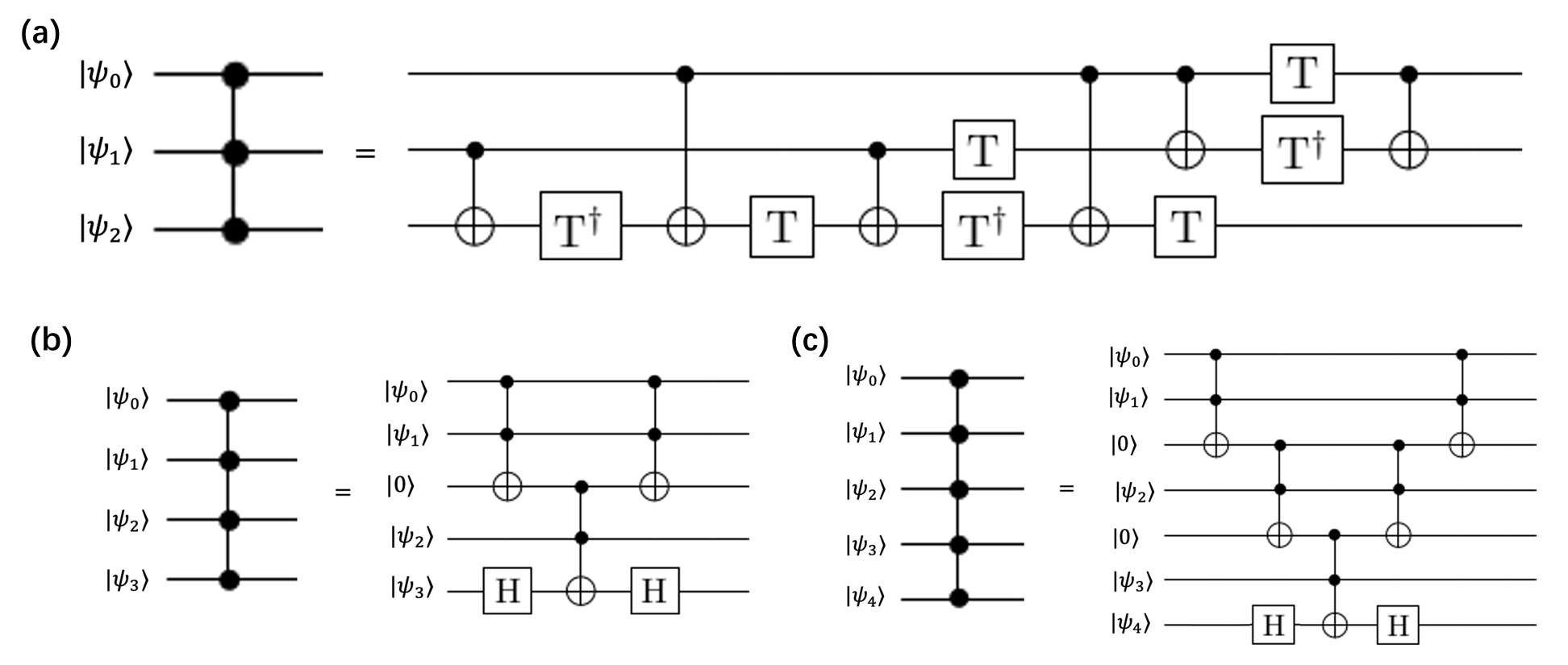}
    \caption{Decomposition of the $C_{N-1}Z$ gate. (a) Implementation of the CCZ gate using CNOT and T gates. (b) Decomposition of $C_{N-1}Z$ gate for $N=4$ using the clean-ancilla V-chain method, which requires ancillary qubits initialized in the state $\ket{0}$. The Toffoli gates in this decomposition can be implemented with CCZ and Hadamard gates, and the CCZ gate can be further decomposed into single-qubit and two-qubit gates using the implementation in (a). (c) Decomposition of the $C_{N-1}$ gate for $N=5$.}
    \label{fig:decomposition_cz}
\end{figure}
\begin{figure}[htbp]
    \centering
    \includegraphics[width=.9\linewidth]{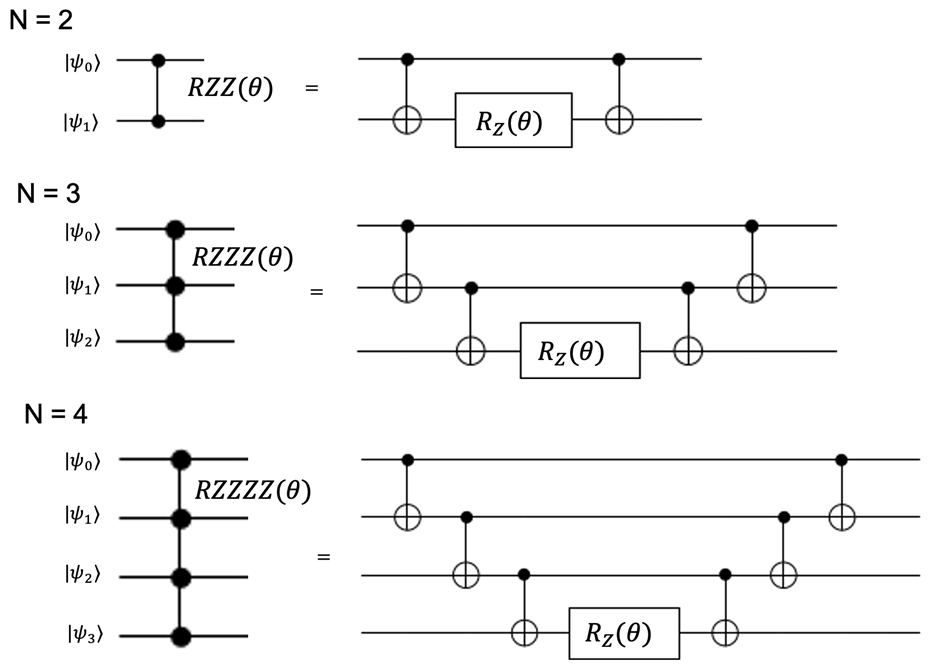}
    \caption{The decomposition of multi-qubit phase rotation gate, for N=2, 3 and 4 respectively.}
    \label{fig:decomposition_rzz}
\end{figure}

As discussed in session IV.C of the main text, directly implementing multi-qubit gates provides significant advantages in both the efficient execution of quantum algorithms and the achievement of high fidelities. In architectures restricted to two-qubit gate operations, multi-controlled gates must be decomposed into sequences of two-qubit gates. For $N=3$, the $CCZ$ gate can be decomposed into a sequence of $CNOT$ gates and $T$ gates, as shown in Fig.~\ref{fig:decomposition_cz}. For $N\geq 4$, $C_{N-1}Z$ gate can be constructed using Hadamard gates and $CCZ$ gates through a V-chain decomposition method~\cite{balauca2022efficient}. Fig.~\ref{fig:decomposition_cz}(b) and (c) illustrate the circuit implementations of the $C_{N-1}Z$ gate for $N=4$ and $N=5$, respectively. Similarly, the decompositions of the multi-qubit phase rotation gate for $N=2,3,4$ are shown in Fig.~\ref{fig:decomposition_rzz}. These decomposed circuits are used to obtain Fig.4 in the main text.

\section{Noise model for single-qubit readout}

This section provides a detailed fidelity analysis of the rapid non-demolition readout scheme discussed in Section V.A of the main text. 

The proposed readout scheme consists of two stages: the entangling stage during which the entanglement is generated between data qubits and atomic ensembles, followed by a detection stage that an EIT-based Rydberg state detection is applied on the atomic ensembles. Errors occurring in the entangling stage propagate to the readout stage, affecting the overall fidelity. Below, we analyze the error mechanisms in details, considering different initial states of the data qubits. 

With this new pulse sequence, when the data qubit is in state $\ket{0}$, it remains unchanged throughout the entire readout process, since it is never coupled to the Rydberg state $\ket{r}$. Consequently, the only source of measurement error is the Rydberg state decay of the atomic ensemble. We assume that the both ensemble states $\ket{A}$ and $\ket{R}$ decay to the ground state $\ket{G}$, which behaves identically to $\ket{A}$ during the later detection stage (\textit{i.e.,} it results in low probe light transmission). This leads to a readout error probability of $p_{m,0}=\Gamma_R(2\pi/\Omega_r+\pi/\Omega_R)$. 


For a data qubit in state $\ket{1}$, the error mechanisms are more complex due to its coupling to $\ket{r}$ during the entangling stage. This coupling is conditioned on the atomic ensemble state via Rydberg interactions. Errors arise from both the spontaneous decay of the data qubit from $\ket{r}$ and the decay of atomic ensemble from state $\ket{R}$ and $\ket{A}$. We first consider the errors caused by the spontaneous decay of the data qubits from $\ket{r}$. During the first $X_{2\pi}$ pulse (see Fig. 5 of the main text), the data qubit spends half of the $2\pi$-pulse duration in $\ket{r}$ when the atomic ensemble is in state $\ket{A}$. The probability of decaying from $\ket{r}$ is $p_{r}=\Gamma_r\pi/2\Omega_r$. In the worst case, this decay prevents the data qubit from returning to $\ket{1}$, instead leaving it outside the computational basis or in $\ket{0}$, both of which cannot be coupled to $\ket{r}$. Consequently, the affected qubit behaves as if it were in $\ket{0}$ in subsequent EIT-based detection stage, effectively introducing a bit-flip error. Additionally, this decay heralds that the atomic ensemble is in $\ket{A}$, since $\ket{1}$ can only be coupled to $\ket{r}$ when atomic ensemble is in $\ket{A}$. After applying the second $X_{\pi/2}$-pulse to the atomic ensemble, it evolves into $(\ket{A}-i\ket{R})/\sqrt{2}$, leading to a measurement error with probability of $p_{m,1}=1/2$ in the EIT-detection stage. Next, we consider the error from the decay of the atomic ensemble Rydberg states. Such decay affects the data qubit's evolution by breaking the Rydberg blockade condition, causing unintended excitation of the data qubit to $\ket{r}$ with probability $\Gamma_R\pi/2\Omega_r$. To restore the population, at the end of the entangling stage, $\ket{r}$ is coupled to a stretched intermediate state $\ket{6s6p^3P_1,F=3/2,m_F=3/2}$, which eventually decays back to $\ket{1}$. In the following EIT-based detection stage, the atomic ensemble Rydberg decay does not directly contribute to measurement errors, as decay back to $\ket{G}$ is indistinguishable from the ensemble remaining in state $\ket{A}$  (\textit{i.e.,} high probe light transmission), which is the expected outcome in the absence of errors. In total, the error during entanglement generation correspond to the Kraus operator
\begin{equation}
\label{readout noise}
\begin{split}
    K_f &= \sqrt{p_{r}}\ket{0}\bra{1} 
\end{split}
\end{equation}
on data qubit, together with readout error with probability $p_{m,0}$ for state $\ket{0}$ and probability $p_{m,1}$ for state $\ket{1}$, which correlates with the occurrence of Kraus operator $K_f$ with probability $p_{r}$. 

\begin{figure}
    \centering
    \includegraphics[width=\linewidth]{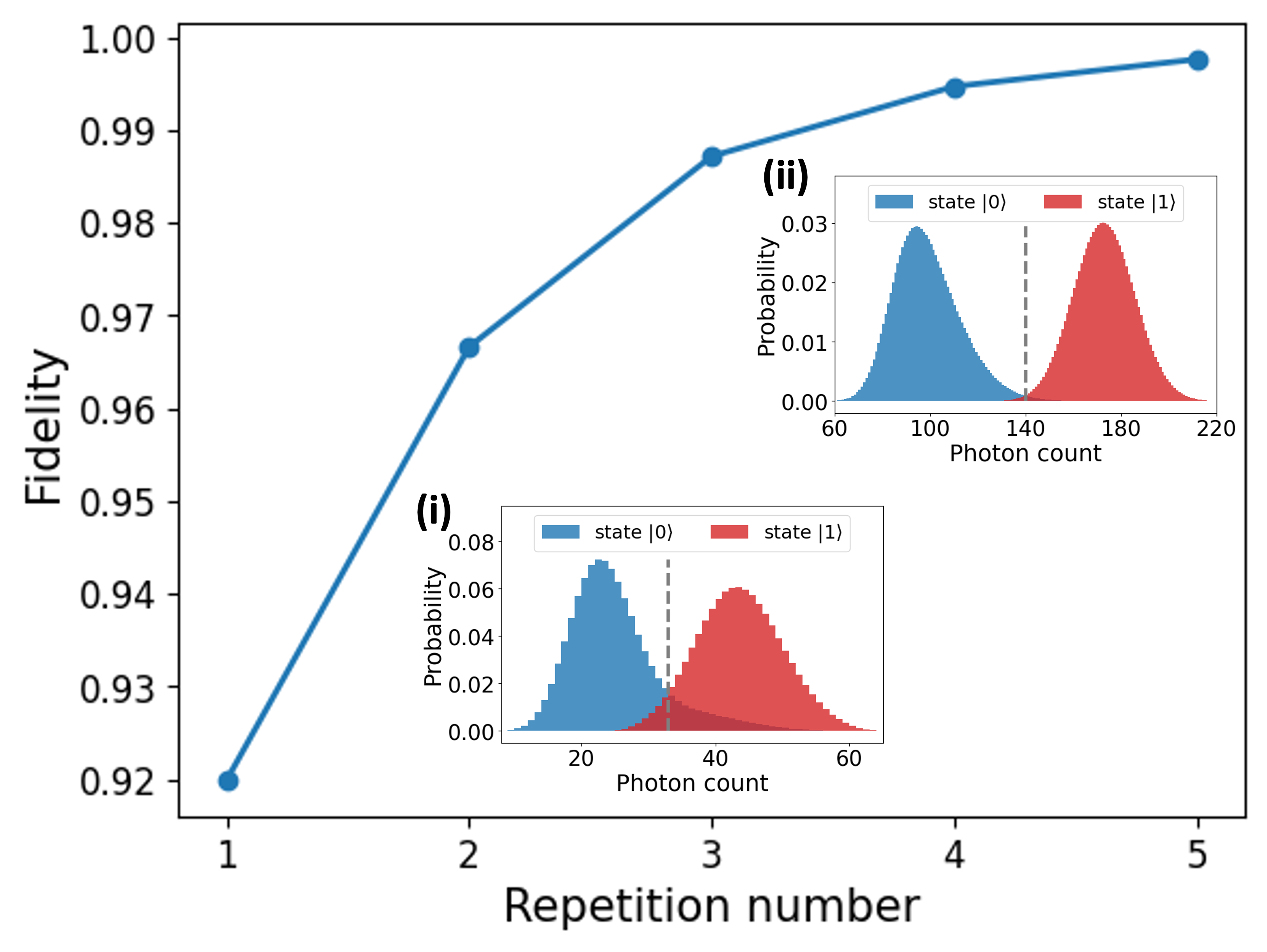}
    \caption{Fidelity of the repetitive readout scheme for different repetition numbers. Starting from a single-shot fidelity of $F_d\approx0.92$, the readout fidelity improves with repetitions, reaching $>99\%$ after four iterations. Inset: Distribution of transmitted probe photon numbers for state detection. (i) Single-shot readout performed over $6\mathrm{\mu s}$. (ii) Repetitive readout with four iterations.}
    \label{fig:readout-fidelity}
\end{figure}

Following entanglement generation, the EIT-based detection scheme from~\cite{xu2021fast} is used to distinguish distinct Rydberg states in the atomic ensembles, equivalent to implementing a projective measurement on the data qubits. This EIT-based detection fidelity depends on the photon count statistics: longer detection durations improve the signal-to-noise ratio (SNR) but limited by the Rydberg state decay. In~\cite{xu2021fast}, the maximum detection time is constrained due to this trade-off, eventually limiting the detection fidelity. In contrast to this previous approach, our proposed scheme, due to its non-demolition nature, allows for multiple measurement rounds to enhance fidelity. To estimate the fidelity after multiple repetitions, we update the data qubit state according to the noise channel (Eq.~\ref{readout noise}) in each round. The total photon count distribution after different repetition numbers is produced by summing the photon counts from each measurement round, accounting that the input state for each new round has been perturbed due to the loss of Rydberg states in the previous rounds. Fig.~\ref{fig:readout-fidelity} plots the photon count histograms for different repetition numbers, using the same experimental parameters and measured Rydberg loss rate from~\cite{xu2021fast}, and $p_{m,0}=7.5\times 10^{-4}$ and $p_{r}=2.5\times 10^{-4}$. It shows the readout fidelity exceeds $F>99\%$ after four repetitions, 

\section{Noise model for multi-qubit joint measurement}
\label{noise model}

In this section, we develop an effective noise model to account for multiple error mechanisms in the multi-qubit measurement scheme, which is an extension based on the single-qubit readout discussion. Since this scheme is particularly relevant for quantum error correction protocols, it is crucial to discuss population leakage due to Rydberg state decay to prevent error propagation across following error correction rounds. 

Rydberg state decay can cause population transfer to nearby Rydberg states via blackbody radiation, or to low-lying states through spontaneous emission, which eventually decay into the ground state or metastable states. Additionally, since the evolution of data qubits is conditioned on the atomic ensemble state via Rydberg interactions, any loss of Rydberg excitations in the atomic ensemble breaks the blockade condition. This can result in unwanted residual population in the Rydberg state $\ket{r}$ after the detection stage. To address population leakage into $\ket{r}$ and unwanted population transfer to other Rydberg or metastable states, we couple the Rydberg state $\ket{r}$ to the stretched intermediate state $\ket{6s6p^{3}P_1,F=3/2,m_F=3/2}$, which decays to the qubit state $\ket{1}$ only. Atoms that populate other Rydberg states can be converted to detectable atom loss using anti-trapping tweezer beams, allowing these atoms to be replaced with fresh ones. Population in metastable states can be restored to the ground-state manifold using incoherent optical pumping~\cite{Cho2012optical}. We assume that the spontaneous decay, the replacement of fresh atoms, and the incoherent pumping leave the atoms in the qubit states $\ket{0}$ and $\ket{1}$ with equal probability. 

Now, we consider a joint measurement on $m$ data qubits using a single atomic ensemble. After addressing population leakage as discussed above, the error due to the decay of the i-th data qubit from $\ket{r}$ results in the following Kraus operators:

\begin{equation}
\begin{split}
    K_{f_i}=\sqrt{\frac{p_r}{2}} I_1\otimes\cdots \ket{0}\bra{1}_i\cdots \otimes I_m\\
    K_{1_i}=\sqrt{\frac{p_r}{2}} I_1\otimes\cdots \ket{1}\bra{1}_i\cdots\otimes I_m
\end{split}
\end{equation}
where $p_r=\Gamma_r\pi/2\Omega_r$ is the decay probability of the data qubit's Rydberg state.

In addition to the data qubit Rydberg decay, the decay of the atomic ensemble Rydberg states also affects the evolution of data qubit states. If decay occurs at time $t$, a data qubit initially in state $\ket{\psi_0}=\alpha\ket{0}+\beta\ket{1}$ evolves as:
\begin{equation}
    \ket{\psi(t)}=\alpha\ket{0}+\beta(-\cos(\frac{\Omega_r t}{2}))\ket{1}-i\sin(\frac{\Omega_r t}{2}\ket{r})
\end{equation}
Averaging over $t$ from $0$ to $t=T=2\pi/\Omega_r$, the resulting density matrix becomes
\begin{equation}
    \rho = \abs{\alpha}^2\ket{0}\bra{0}+\frac{1}{2}\abs{\beta}^2(\ket{1}\bra{1}+\ket{r}\bra{r})
\end{equation}
After restoring the population in $\ket{r}$ back to $\ket{1}$, the density matrix becomes
\begin{equation}
    \rho = \abs{\alpha}^2\ket{0}\bra{0}+\abs{\beta}^2\ket{1}\bra{1}=M_0\rho_0M_0^{\dagger}+M_1\rho_0M_1^{\dagger}
\end{equation}
where $\rho_0=\ket{\psi_0}\bra{\psi_0}$, $M_0=\ket{0}\bra{0}$, and $M_1=\ket{1}\bra{1}$. Considering this error to all data qubits, this process corresponds to the Kraus operators:
\begin{equation}    
K_{\mathbf{s}}=\sqrt{p_R}\ket{\mathbf{s}}\bra{\mathbf{s}}
\end{equation}
where $p_R=\Gamma_R\pi/\Omega_r$ is the decay probability of $\ket{R}$ and $\ket{\mathbf{s}}=\ket{s_1,s_2,\cdots,s_m}$ with $s_i=0$ or $1$ for $i=1,\cdots,m$.

Using the Pauli twirling approximation and the identities $\ket{0}\bra{0}=(I+Z)/2$, $\ket{1}\bra{1}=(I-Z)/2$ and $\ket{0}\bra{1}=(X+iY)/2$, we have:
\begin{equation}
    \begin{split}
        K_{f_i}\rho_d K_{f_i}^{\dagger}&=\frac{p_r}{8}(X_i\rho_dX_i+Y_i\rho_d Y_i)\\
        K_{1_i}\rho_d K_{1_i}^{\dagger}&=\frac{p_r}{8}(I_i\rho_d I_i+Z_i\rho_d Z_i)\\
        K_{\mathbf{s}}\rho_d K_{\mathbf{s}}^{\dagger}&=\frac{p_R}{2^{2m}}\sum_{P\in \{I,Z\}^{\otimes m}}P\rho_d P^{\dagger}
    \end{split}
\end{equation}
which results in an effective quantum channel on data qubits as:
\begin{equation}
    \label{noise}
    \begin{split}
        \mathcal{E}(\rho_d)=&\left(1-\frac{3m}{8}p_r-\frac{2^{m}-1}{2^{m}}p_R\right)\rho_d\\
        +&\frac{p_r}{8}\sum_{i=1}^m(Z_i\rho_d Z_i+X_i\rho_d X_i+Y_i\rho_d Y_i)\\
        +&\frac{p_R}{2^{m}}\sum_{P\in \{I,Z\}^{\otimes m}\atop P\neq I^{\otimes m}} P\rho_d P^{\dagger}.
    \end{split}
\end{equation}
During the following EIT-based detection stage, the probability of incorrectly associating the transmitted light intensity with the correct joint measurement outcome is given by $p_{m}=p_R/2+mp_r/2+p_{m,d}$. The first two terms arise from decay events occuring during the entangling stage, which cause the atomic ensemble to end in an incorrect state. The term $p_{m,d}$ accounts for the infidelity of the EIT-based detection stage. This physical noise model is used to evaluate the performance of error syndrome measurements for the standard surface code in Section VI.A of the main text.

\newpage
\bibliography{supplement_ref}